\pdfoutput=1
\documentclass[12pt,aps,tightenlines,showkeys]{revtex4}
\usepackage{amsmath}
\usepackage{amssymb}
\usepackage{bm}
\usepackage[mathscr]{eucal}
\usepackage{theorem}
\usepackage{graphicx}
\usepackage{psfrag}
\usepackage{subfigure}
\usepackage{color}
\include{graphix}

\numberwithin{equation}{section}
\graphicspath{{figs/}}
\newtheorem{prop}{Proposition}

\newcommand{\savecomment}[1]{}

\newcommand{\mathnotation}[2]{\newcommand{#1}{\ensuremath{#2}}}

\mathnotation{\pd}{\partial}                    
\mathnotation{\td}{\bm{d}}                      
\mathnotation{\md}{\bm{D}}
\mathnotation{\ee}{{\mathrm e}}                 
\mathnotation{\imi}{\mathrm{i}}                 
\mathnotation{\ldef}{\mathrel{\raisebox{.069ex}{:}\!\!=}}
\mathnotation{\rdef}{\mathrel{=\!\!\raisebox{.069ex}{:}}}
\mathnotation{\dint}{\,{\mathrm{d}}}            
\mathnotation{\com}{\,,\,}

\mathnotation{\grad}{\nabla}                    
\mathnotation{\curl}{\grad\times}               
\mathnotation{\lapl}{\Delta}                  
\mathnotation{\Connection}{\Gamma}                 

\mathnotation{\J}{J}

\mathnotation{\Sub}{\mathcal{S}}                 
\mathnotation{\Free}{\mathcal{S}^{*}}            
\mathnotation{\xv}{\bm{x}}     
\mathnotation{\qv}{\bm{q}}                    
\mathnotation{\Qv}{\bm{Q}}                    
\mathnotation{\yv}{\bm{y}}
\mathnotation{\x}{x}                             
\mathnotation{\pres}{p}                          
\mathnotation{\uv}{\bm{u}}                       
\mathnotation{\uc}{u}                            
\mathnotation{\Xv}{\bm{X}}                       
\mathnotation{\X}{X}                             
\mathnotation{\Yv}{\bm{Y}}                       
\mathnotation{\Y}{Y}                             
\mathnotation{\Vv}{\bm{V}}                       
\mathnotation{\V}{V}                             
\mathnotation{\Metricc}{\mathbb{G}}              
\mathnotation{\Metricm}{[\Metricc]}              
\mathnotation{\metricc}{\mathfrak{g}}            
\mathnotation{\metricm}{[\metricc]}              
\mathnotation{\normalv}{\bm{n}}                  
\mathnotation{\normal}{n}                        
\mathnotation{\Pv}{\bm{P}}                       
\mathnotation{\Pc}{P}                            
\mathnotation{\s}{s}                             
\mathnotation{\F}{F}                             
\mathnotation{\G}{G}                             
\mathnotation{\z}{z}                             
\mathnotation{\Rc}{R}                            
\mathnotation{\T}{T}                             
\mathnotation{\Hc}{H}                            
\mathnotation{\Sc}{S}                            

\begin{document}
\title{Chemical reactions in the presence of surface modulation and stirring}
\author{Khalid Kamhawi$^{1}$ and Lennon \'O N\'araigh$^{2}$}
\email{lennon.o-naraigh05@imperial.ac.uk}
\affiliation{Department of Mathematics$^{1}$ and Chemical Engineering$^{2}$, Imperial College
London, SW7 2AZ, United Kingdom}
\date{\today}

\begin{abstract}
We study the dynamics of simple reactions where the chemical species
are confined on a general, time-modulated surface, and subjected to externally-imposed
stirring.  The study of these inhomogeneous effects requires a model based on a reaction-advection-diffusion
equation, which we derive.  We use homogenization methods to
show that up to second order in a small scaling parameter, the modulation
effects on the concentration field are asymptotically equivalent for systems with
or without stirring.  This justifies our consideration of  the simpler reaction-diffusion
model, where we find that by  modulating the substrate, we can modify the
reaction rate, the total yield from the reaction, and the speed of front
propagation.  These observations are confirmed in three numerical case studies
involving the autocatalytic and bistable reactions on the torus and a sinusoidally-modulated
substrate.
\vskip 0.1in
\noindent{EES classification}: 100.000, 110.000
\end{abstract}

\keywords{Fisher-KPP equation; Advection; Homogenization theory; Manifolds}

\maketitle


\section{Introduction}
\label{sec:intro}
We investigate the dynamics of the logistic and bistable reactions on a generalized,
time-varying substrate with stirring.  Such spatially-inhomogeneous problems
call for the solution of a reaction-advection-diffusion equation, and much
chemical and biological activity in fluid flow can be modelled by such equations.
 In particular, problems concerning autocatalytic chemical reactions~\cite{Strogatz1994}
 and population dynamics~\cite{Skellam1951,Murray1993}
  possess a logistic growth function as a reaction term, and thus satisfy
  a Fisher-KPP type of equation.  Other, more complicated growth functions
  can be used to model a variety of phenomena, including the spread of insect
  populations, or the propagation of electro-chemical waves in organisms~\cite{Murray1993}.
  The original motivation of this
  study was to understand the effects of wave modulation on the population
  dynamics of plankton, although the language and techniques we use are quite
  general.

Before deriving and analyzing our model, we place our work in context by
examining several streams of work that are relevant.
It is known that a growing domain can modify biological pattern
formation, as evidenced by the work of Newman and Frisch~\cite{Newman1979}.
 This has given impetus to the study of reaction-diffusion equations on growing
 domains~\cite{Kondo1995,Crampin1999} in one dimension.  Logically, this
 has led to the study of such problems on manifolds embedded in three dimensions.
  In multiple dimensions, one considers either the effect of curvature~\cite{Gomatam1997,Varea1999,Chaplain2001},
  or the twin effects of domain growth and curvature~\cite{Plaza2004,Gjorgjieva2007}.
  The paper of Plaza \textit{et al.}~\cite{Plaza2004} is particularly relevant
  to the
  present work.  In it, the authors derive the reaction-diffusion equation
  for a class of manifolds, and then study pattern formation on growing domains.
   We re-work their derivation to include the most general two-dimensional
   (differentiable) manifold possible, and then shift the focus from pattern
   formation to
   reactions in the presence of stirring.  The geometric formalism of Aris~\cite{Aris1962}
   is central to our derivation.  These references~\cite{Plaza2004,Gjorgjieva2007}
   consider the `geometric sink', that is, the notion that a growing domain
   can act as a sink for the chemical reaction.  We extend this idea to growing
   and modulating domain sizes and examine the effects of the sink through
   numerical simulation.
   
The notion of flow-driven reactions is not new.  In~\cite{Neufeld2001}, the
effects of chaotic advection on the Fitz Hugh--Nagumo model are considered.
 The flow is found to produce a coherent global excitation of the system,
 for a certain range of stirring rates.  The effects of flow can also induce
 distinctive spatial structure in the chemical concentration; this is studied
 in~\cite{Neufeld1999}.  The same authors examine the single-component logistic
 or Fisher-KPP model in~\cite{Neufeld2001}.  There the focus is on regime-change, namely
 how the rate of chaotic advection affects the spatial structure of the concentration.
  For slow stirring / fast reactions, a spatially inhomogeneous perturbation
  decays rapidly, and the equilibrium state is reached rapidly.  On the other
  hand, for fast
  stirring / slow reactions, the perturbation persists, and a
  filament structure propagates throughout the domain.  Nevertheless, the
  asymptotic state is still a stable homogeneous one.  This system is simplified
  and the the transition treated as a bifurcation problem in~\cite{Menon2005}.
   The focus of these papers is on local temporal and spatial structure.
    It is therefore salutary to examine the paper of Birch \textit{et al.}~\cite{Birch2007},
    where the authors examine the averaged effect of a non-constant growth
    rate on the dynamics of the stirred Fisher-KPP equation.  The authors use
    the theory of estimates to obtain bounds on the reaction yield, as a
    function of the stirring and the non-constant growth rate.  When the
    mean growth rate is negative, the previously-unstable zero state of the
    Fisher-KPP equation can become stable; then the catalyst fails to propagate the
    reaction.  In the Birch paper, the inhomogeneous growth rate is the consequence
    of an inhomogeneous distribution of nutrient in a plankton population.
     It could, however, be the result of placing the population or chemical
     species on a modulating surface, which is the subject of our report.
      Indeed, our results demonstrate the possibility of increasing the reaction
      yield by surface modulation.

If the flow field or the modulation have small length scales compared to
the domain size, then homogenization theory naturally presents itself
as a tool for understanding the effects of flow and modulation in an averaged
sense~\cite{Pavliotis2008}.  The small scales are bundled up into an effective
diffusion constant, and the model reduces to a more manageable equation involving
a diffusion operator.  Such an approach has been used in the linear advection-diffusion
equation~\cite{McLaughlin1985,McCarty1988,Rosencrans1997}, for linear reaction-advection-diffusion equations~\cite{Papanicolaou95},
and for non-linear equations~\cite{Bensoussan1978}.  We propose and justify the application of this theory
to a linearized, advective Fisher-KPP equation on a time-varying manifold, and
compute the effective diffusivity for the torus.  The effective diffusivity
for a manifold is, in general, a function of position, although in the fortuitous
case we consider, this dependence vanishes.  As in~\cite{Neufeld2002},
the asymptotic state of the system is homogeneous, and the purpose of our
 homogenization calculation is therefore to study the speed at which this state is reached.

This paper is organized as follows.  In Sec.~\ref{sec:background} we discuss
the autocatalytic reaction in the homogeneous case, and then generalize this
formalism to consider reactions on time-dependent, spatially-inhomogeneous surfaces.  In Sec.~\ref{sec:homo} we derive conditions on the metric tensor
for homogeneous solutions to exist on a general surface.  We then outline
a separation-of-scales technique that enables us to compute the spatial distribution
of concentration as the solution of a diffusion equation.  We discuss the
case of shear flow on a modulating torus, where the equation for the effective
diffusion is remarkably simple.  In Sec.~\ref{sec:num} we outline three case
studies for unstirred mixtures.  We demonstrate that modulating the surface
can increase the reaction yield.  We verify that this result is independent
of the reaction type by obtaining a similar result for the bistable reaction
function.  Finally, in Sec.~\ref{sec:conc} we present our conclusions.

\section{Theoretical formulation}
\label{sec:background}

In this section we describe the mathematical formulation for the autocatalytic
reaction.  We derive the rate equation for homogeneous concentrations, and
then generalize this result to spatially-varying concentrations on generalized
two-dimensional surfaces.  We pay close attention to understanding the effect
on the reaction rate when the surface itself varies with time.

In the homogeneous case, the evolution of $n$ chemical species whose concentrations
are given by the
 vector $\bm{C}=(c_{1}(t),c_{2}(t), ..., c_{n}(t))$
 can be placed in the following canonical form:
\begin{equation}
\frac{d \bm{C}}{d t} = \bm{F}\left(\bm{C}\right),
\label{r-a-d}
\end{equation}
where the $n$-vector of
functions $\bm{F}$ is the reaction term describing the interactions between
the different species.  We are interested in particular in the autocatalytic
reaction $c_{1}+c_{2}\longrightarrow 2c_{2}$; 
the dynamical system for such a reaction is given by the equation pair
\begin{equation}
\frac{d c_{1}}{d t} = -\beta c_{1}c_{2},\; \frac{d c_{2}}{d t} = \beta c_{1}c_{2},
\end{equation}
where $\beta>0$ is the reaction rate.
The implied equation $d(c_{1}+c_{2})/{d t}=0$ is a statement
of molecular conservation. This system is reduced to a single equation
by defining a new variable $c =c_{2}/\left(c_{1}+c_{2}\right)$, giving rise
to the logistic growth law
\begin{equation}
\frac{d c}{d t} = \sigma c(1-c),
\label{FastReaction}
\end{equation} 
where $F(c)= \sigma c(1-c)$ is the reaction function and
$\sigma\equiv\beta(c_{1}+c_{2})>0$ is the associated rate. 
The evolution of this relative concentration is a contest between
linear creation and quadratic destruction, which manifests itself through
the sigmoid solution $c= c(0) e^{\sigma t}/\left[1+c(0) (e^{\sigma t}-1)\right]$,
where $c(0)>0$ is the initial concentration. 

The two critical points satisfying ${d c}/{d t}=0$ are the states $c=0$ and $c=1$, respectively indicating when species
$c_{2}$ is extinct and when the whole space of concentrations is equal to
that of the product $c_{2}$; i.e. $c_{1}$ is extinct. Since ${\partial F}/{\partial
c}=\sigma (1-2c)$,
$c=0$ is unstable and $c=1$ is stable.  
The phase portrait of the
one-dimensional dynamical system is easily envisioned: there is a quadratically-increasing reaction away from the repeller $c=0$, towards larger values of
$c$.   When the product is half of the whole concentration mix $c=1/2$;
the reaction rate is maximal, and thereafter it decreasess
quadratically  towards the attractor at $c=1$.  Hence, if we start with
a soup composed of only $c_{1}$-molecules, the addition of any amount of the second species (no matter how small) leads to the annihilation of the first, while in the reversed scenario it is what is added, in this case
the $c_{1}$-molecules, that will be destroyed. Any mixture of the two tends
towards a homogeneous soup wholly composed of the species $c_{2}$.
%
%
%
 
We can generalize
the mass-action law to the inhomogeneous case by using the following assumptions:
\begin{itemize}
\item There is diffusion of concentration, arising from thermodynamic fluctuations;
\item There is a large-scale imposed stirring, modelled as an advecting velocity;
\item The substrate $\mathcal{M}$ on which the substance is placed is modulated as
a function of time.
\end{itemize}
We write down a continuity equation that takes account of these features.
 The approach we take was discussed by Aris~\cite{Aris1962}, although the application
 to reaction-diffusion systems is new.
 We examine the mass balance for the chemical concentration with respect
 to a control area $S\left(t\right)$.  We work with a general two-dimensional
 manifold $\mathcal{M}$.  At time $t=0$ the manifold is endowed with co-ordinates
 $\bm{a}_0$, such that
\[
\mathcal{M}\left(0\right)=\{\bm{x}\in\mathbb{R}^3|\bm{x}=\bm{x}\left(a_0^1,a_0^2\right)\},
\]
These co-ordinates can be used to label the fluid particles at time $t=0$.
 As time evolves, the fluid particles are advected by an imposed flow $\bm{U}$,
 and the particle labels develop time-dependence, $\bm{a}\left(t\right)$.
We introduce another set of co-ordinates denoted by $\qv\left(t\right)$.
 These are fixed in the sense that a point in $\mathbb{R}^3$,
 if located by the co-ordinates $\qv\left(t\right)$, has an instantaneous
 velocity wholly normal to the surface $\mathcal{M}\left(t\right)$.  Since the manifold
 varies smoothly in time, there is a set of transformations connecting these
 co-ordinate systems:
\begin{equation}
\qv=\qv\left(\bm{a},t\right),\qquad
\bm{a}=\bm{a}\left(\qv,t\right),\qquad
\bm{a}_0=\bm{a}\left(\qv\left(0\right),0\right).
\label{eq:transf}
\end{equation}
A particle that is advected from an initial point $\bm{a}_0$ by the imposed
stirring $\bm{U}$ therefore has a velocity
\[
\bm{U}\left(\qv,t\right)=\left(\frac{\partial\qv}{\partial t}\right)_{\bm{a}}\equiv\frac{d\qv}{dt},
\]
where, by Eq.~\eqref{eq:transf}, the time derivative is taken at fixed particle
label $\bm{a}$.
The last piece of formalism needed is the prescription of a metric tensor:
\begin{equation}
g_{ij}\left(\qv,t\right)=\frac{\partial\bm{x}}{\partial q^i}\cdot\frac{\partial\bm{x}}{\partial
q^i}=\bm{g}_{(i)}\cdot\bm{g}_{(j)},\qquad g=\mathrm{det}\left(g_{ij}\right),
\label{eq:metric_tensor}
\end{equation}
and the fixedness of the co-ordinate system $\qv$ is thus made manifest:
\[
\bm{g}_{(i)}\cdot\left(\frac{\partial\bm{x}}{\partial t}\right)_{\qv}\equiv\,\,\bm{g}_{(i)}\cdot\frac{\partial\bm{x}}{\partial t}=0,\qquad i=1,2.
\]
Using the transformations~\eqref{eq:transf} and the metric tensor~\eqref{eq:metric_tensor},
we obtain a convenient definition of area, either as an integral over
a fixed domain, or a time-varying one:
\[
\int_{S\left(t\right)}dS=\int_{S\left(t\right)}\sqrt{|g|}dq^1dq^2=\int_{S\left(0\right)}\sqrt{|g|}
J da^1 da^2,
\] 
where $S\left(0\right)$ is the pre-advected domain and
\[
J=\frac{\partial\left(q^1,q^2\right)}{\partial\left(a^1,a^2\right)}
\]
is the Jacobian of the transformation.
This formalism facilitates the derivation of an analog of the Reynolds transport
theorem for a concentration field $c\left(\qv\left(t\right),t\right)$~\cite{Aris1962,Bicak1999,Hu2007}:
\begin{eqnarray}
\frac{d}{dt}\int_{S\left(t\right)}c\,dS&=&\int_{S\left(t\right)}\left[\left(\frac{\partial
c}{\partial t}\right)_{\bm{a}}+c\left(\frac{\partial}{\partial t}\log\sqrt{|g|}\right)_{\bm{a}}\right]dS,\nonumber\\
&=&\int_{S\left(t\right)}\left[\left(\frac{\partial
c}{\partial t}\right)_{\bm{q}}+\mathrm{div}\left(\bm{U}c\right)+c\left(\frac{\partial}{\partial
t}\log\sqrt{|g|}\right)_{\qv}\right]dS.
\label{eq:ddt}
\end{eqnarray}
%
%
%
%
%
%
%
This change in the amount of concentration in the control patch must be matched
by the diffusive flux through the boundary of the patch, and by the amount
\begin{figure}[htb]
\centering\noindent
\includegraphics[width=0.35\textwidth]{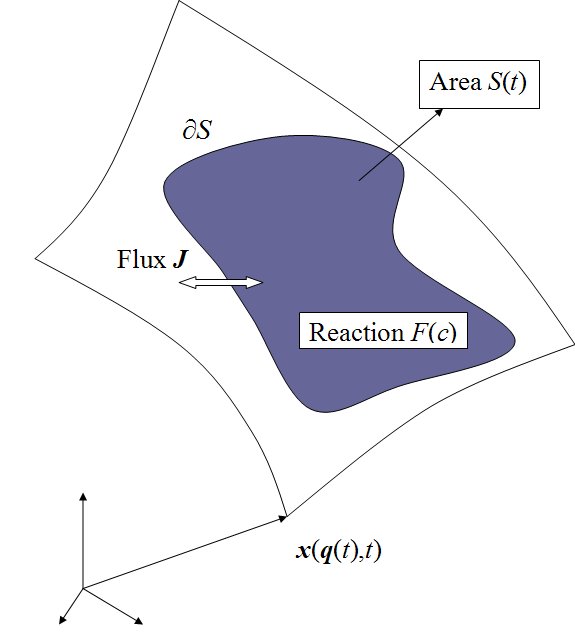}
\caption{A schematic diagram showing the flux of matter into, and out of
a patch of area $S$.  This flux comprises advective and diffusive parts,
and is given by $\bm{J}=\bm{U}c-\kappa\,\mathrm{grad}c$, and $d\bm{\ell}$
is a line element on the curve formed by the boundary of the area $S$.}
\label{fig:geometry}
\end{figure}
of matter created or destroyed by the reaction (shown schematically in Fig.~\ref{fig:geometry}), that is,
\[
\frac{d}{dt}\int_{S\left(t\right)}c\left(\qv,t\right)dS=-\int_{\partial S\left(t\right)}\kappa\,\mathrm{grad}\,c\cdot
d\bm{\ell}+\int_{S\left(t\right)}F\left(c\right)dS,
\]
where $\kappa$ is the (constant) diffusion coefficient and $\partial S$ is the
boundary of $S$.  A simple application
of Gauss's law then gives
\begin{equation}
\frac{d}{dt}\int_{S\left(t\right)}c\left(\qv,t\right)dS=\int_{S\left(t\right)}\kappa\lapl
c dS+\int_{S\left(t\right)}F\left(c\right)dS,\qquad \lapl=\mathrm{div}\,\mathrm{grad}.
\label{eq:cons2}
\end{equation}
Combining Eqs.~\eqref{eq:ddt} and~\eqref{eq:cons2} gives the following
local conservation law:
\begin{equation}
\frac{\partial c}{\partial t}+\mathrm{div}\left(\bm{U}c\right)=\kappa\lapl c+F\left(c\right)-c\frac{\partial \sqrt{|g|}}{\partial t}.
\label{eq:c_space}
\end{equation}
The divergence term $\mathrm{div}\left(\bm{U}c\right)$ can be re-written
as $U^i\partial_{q_i}c=\bm{U}\cdot\nabla_q c$ for incompressible flows. 
We call the term $-c\left(\partial \log\sqrt{g}/\partial t\right)$ the \textit{geometric
sink}: its inclusion is necessary to conserve the total number of particles
on a time-varying substrate.  Note that in  previous applications~\cite{Gjorgjieva2007,Plaza2004}
the scale function
was taken to be a growing function of time, and hence this extra term was
indeed a sink; here we consider a general growth function, and thus this
term can also act as a source.
The geometric source / sink has the following interpretation:
given a concentration equation for the number of
particles per
unit volume, a local source can be introduced in two ways.  The first and
more obvious way is to inject particles into the system.  Here, instead of
increasing or decreasing the local number of particles, we stretch or squeeze
the local area element, so that the local concentration changes.  This
effect vanishes upon integration, so that the total number of particles is
conserved in a global sense, although locally, the number of particles per
unit volume changes because the volume itself changes.

We non-dimensionalize Eq.~\eqref{eq:c_space}
to understand the relative importance of stirring, diffusion and reaction
kinetics on the dynamics. Given a characteristic length scale $L$, and a
characteristic speed $U$ of an incompressible velocity field $\bm{U}$, Eq.~\eqref{eq:c_space}
is parametrised by the P\'eclet number $Pe={L U}/{\kappa}$ and the Damk\"ohler
number $Da={\sigma L}/{U}$, such that
\begin{equation}
\frac{\partial c}{\partial t}+\mathrm{div}\left(\bm{U}c\right)
= Pe^{-1}\Delta c + Da\;c(1-c)-c\frac{\partial\mathrm{log}\sqrt{|g|}}{\partial t};
\label{eq:nondim_rad}
\end{equation}
these groups respectively describe the ratio of 
the advective/diffusive and chemical/advective timescales.
For any stirring variation, the group $Da Pe = \sigma L^2/\kappa=Const.$ is unchanged.  Hece for a fixed diffusion coefficient $\kappa$, the chemical
timescale $1/\sigma$ is the only free parameter controlling
the dynamics.  Thus, by keeping the other parameters fixed, the chemical
timescale sufficiently represents the dynamical timescale of the
reaction-advection-diffusion equation~\cite{Neufeld2001}. 
Accordingly, the case of a homogeneous concentration field discussed above
\eqref{FastReaction} is equivalent to having a very fast reaction rate with respect to a fixed diffusion of order $O(Pe)=1$
and no advection, i.e. $Da\gg 1$. From now on we shall fix the diffusion
rate at order unity and describe the dynamics in similar terminology, such
that the reaction takes place at a rate measured against the diffusion rate.


\section{Scale separation with stirring and surface modulation}
\label{sec:homo}

The macroscopic behaviour of a system with phenomena occurring at
various length and time scales can be described by homogenization theory.
The PDE (and its boundary conditions) that describes the system, is analyzed
 as having rapidly oscillating differential operators corresponding
to the different scales of the phenomena. Taking the appropriate limit of
infinite
scale separation, the solution of the homogenized PDE (known as the cell
problem) describes the large-scale behavior induced by the small-scale dynamics.
 In this section, we use this approach to calculate the effective diffusivity
 for the reaction-advection-diffusion equation, for a certain class of substrates.
  The type of substrate modulation we specify is rather restrictive, although
  our results will demonstrate the qualitative effects of substrate modulation.

\subsection{The uniform solution}

If the metric tensor in the reaction-advection-diffusion equation~\eqref{eq:nondim_rad}
satisfies certain restrictions, a homogeneous concentration field $c_{0}(t)$
exists.  To see this, we set the operators $\mathrm{div}$ and $\lapl$ to
zero in the equation, which gives rise to the following form:
\begin{equation}
\frac{dc_0}{dt}=Da \;c_0\left(1-c_0\right)
-c_0\frac{\partial\log \sqrt{g}}{\partial t}.
\label{eq:Bernoulli}
\end{equation}
If, either
\begin{itemize}
\item $\sqrt{|g|}=\rho\left(t\right)G\left(\bm{q}\right)$, or more restrictively,
\item $g_{ij}=\rho\left(t\right)G_{ij}\left(\bm{q}\right)$,
\end{itemize}
then we call the metric tensor separable or fully separable respectively.
 We call the time-dependant separable term $\rho(t)$
the \textit{scale function}.  Note the following observation:
\[
\mathrm{det}({g}_{ij})=\rho G \text{ does not imply that } {g}_{ij}=\rho {G}_{ij}.
\]
One example of such non-equality is furnished by the metric of the torus
in Eq.~\eqref{eq:TorusMetric}.  By choosing an appropriate modulation of the
inner and outer toroidal radii, the metric of the torus can, however, be
made fully separable.
Thus, using a separable metric, we have the equality
\begin{equation}
\frac{\partial\log \sqrt{g}\left(\qv,t\right)}{\partial t} 
= \frac{1}{\rho(t)}\frac{d \rho(t)}{d t} 
\label{separablemetric}
\end{equation}
and hence the differential equation for the uniform concentration is itself
uniform, and this approach is self-consistent.
Restricting to this class of substrate modulation, the homogeneous
concentration field has an explicit solution as the solution of a Bernoulli
equation~\cite{Polyanin_odes}:
\begin{equation}
c_0\left(t\right)=\frac{e^{\int_{0}^{t} (Da-\frac{d\log{\rho}}{dt'})dt'}}
{\frac{1}{c_{0}(0)}+Da\;\int_{0}^{t}\{ e^{\int_{0}^{t'} (Da-\frac{d\log{\rho}}{dt''})dt''}\}dt'}.
\label{bernoulli_sln}
\end{equation}
In the following applications, we shall make use of the \textit{growth function}
\[
\gamma\left(t\right)=Da\left[t-2\int_0^tc_0\left(s\right)ds\right],
\]
which satisfies the following important result:
\begin{prop}[The growth function $\gamma\left(t\right)$ is bounded in time.]\label{prop:growth}
This result holds when the scale function $\rho\left(t\right)$ is bounded
below, $0<\rho_{\mathrm{min}}\leq \rho\left(t\right)$.   We re-write
$c_0\left(t\right)$ as a perfect derivative, using the notation $f(t)\ldef\rho^{-1}(t)e^{Da\;t}$
\[
c_{0}(t)=\frac{1}{Da}\frac{d\log{(
\frac{1}{c_0(0)}+Da\int_{0}^{t}f(s)ds)}}{dt},
\]
so that its integral is simply
\begin{eqnarray*}
\int_0^tc_{0}(s)ds
&=&\frac{1}{Da}\bigg\{\log\left[
\frac{1}{c_0(0)}+Da\int_{0}^{t}f(s)ds\right]-\log\left(\frac{1}{c_0(0)}\right)\bigg\},\\
&=&\frac{1}{Da}\log\left[1+Da c_0\left(0\right)\int_{0}^{t}f(s)ds\right].
\end{eqnarray*}
Thus,
\begin{eqnarray*}
\gamma\left(t\right)&=&Da\,t-2\log\left[1+Da c_0\left(0\right)\int_{0}^{t}f(s)ds\right],\\
&=&2\bigg\{\log\left(e^{{Da\,t}/2}\right)-\log\left[1+Da c_0\left(0\right)\int_{0}^{t}f(s)ds\right]\bigg\},\\
&=&2\log\left[\frac{e^{{Da\,t}/2}}{1+Da c_0\left(0\right)\int_{0}^{t}f(s)ds}\right],\\
&\leq&2\log\left[\frac{e^{{Da\,t}/2}}{1+\frac{Da c_0\left(0\right)}{\rho_{\mathrm{min}}}\int_{0}^{t}e^{Das}ds}\right],
\end{eqnarray*}
and this last quantity has a $t$-independent upper bound, $\gamma_0$.
\end{prop}

\subsection{The homogenized solution}
We homogenize the advection-reaction-diffusion equation on 
a manifold $\mathcal{M}$.  For this approach to work, we must specialize to a manifolds
with certain special properties.
 Because homogenization theory is, in general, applicable only to periodic
 domains, the surface $\mathcal{M}$ must either be a periodic surface embedded in $\mathbb{R}^3$,
 or the torus $\mathbb{T}^2$, with an appropriate modulation of the radii.  We make the further restriction that the metric
 tensor of the manifold $\mathcal{M}$ be fully separable, and that the modulation
 of the manifold is periodic in time.  The separability condition means that
 the co-ordinates $\qv$ are independent of time.
Then, the advection-reaction-diffusion equation, written in co-ordinate form
$\bm{q}=\left(q^1,q^2\right)$, is the following:
\[
\frac{\partial c}{\partial t}+\frac{1}{\sqrt{|G|}}\frac{\partial}{\partial q^i}\left(\sqrt{|G|}U^ic\right)=\frac{Pe^{-1}}{\rho\left(t\right)\sqrt{|G|}}\frac{\partial}{\partial
q^i}\left(\sqrt{|G|}G^{ij}\frac{\partial c}{\partial q^j}\right)+F\left(c\right)-c\frac{\partial\log\rho}{\partial
t}.
\] 
Thus, we have isolated the modulation terms, and the equation can be expressed
in a form where the differential operators are independent of time: 
\begin{equation}
\frac{\partial c}{\partial t}+\mathrm{div}_0\left(\bm{U}\left(\bm{x},t\right)c\right)=\frac{\rho\left(0\right)}{\rho\left(t\right)}Pe^{-1}\lapl_0c+F\left(c\right)-c\frac{\partial\log\rho}{\partial
t},
\label{eq:rad_man}
\end{equation}
where $\lapl_0\phi=\rho\left(0\right)^{-1}|G|^{-1/2}\partial_{q_i}\left(|G|^{1/2}G^{ij}\partial_{q_j}\phi\right)$.
 For ease of notation, we absorb the prefactor $\rho\left(0\right)$ into
 the inverse P\'eclet number. 

Solving the full logistic model \eqref{eq:rad_man} is problematic, as
it is a non-linear equation.  Fortunately, it is possible to understand the
distribution of spatial variations in detail simply by studying the linearized
form
\begin{equation}
c(\qv,t)=c_{0}(t)+\delta \psi(\qv,t),
\label{eq:cgensol}
\end{equation} 
where $\delta\ll1$
and the advection only affects the second term, $\psi\left(\qv,t\right)$. Then, the reaction-advection-diffusion equation becomes
\begin{equation}
\left(\frac{\partial}{\partial{t}}-Da\left[1-2c_0\left(t\right)\right]+\frac{\partial\log\rho}{\partial
t}\right)\psi=
-\mathrm{div}_0\left(\bm{U}\psi\right) + \frac{Pe^{-1}}{\rho\left(t\right)}\Delta_0\psi.\qquad
\label{eq:deltac2}
\end{equation}
This evolution equation provides a uniform bound on $\|\psi\|_2^2$, and hence
the decomposition~\eqref{eq:cgensol} is valid for all times.
\begin{prop} [The quantity $\|\psi\|_2$ is uniformly bounded.]\label{prop:bdd}  To see this,
multiply Eq.~\eqref{eq:deltac2} by $\psi\sqrt{|g|}dq^1dq^2$ and integrate.
 Using the incompressibility condition, this gives the equation
\[
\frac{d}{dt}\|\psi\|_2^2=-\frac{Pe^{-1}}{\rho\left(t\right)}\|\mathrm{grad}\psi\|_2^2+\underbrace{Da\left(1-2c_0\left(t\right)\right)}_{=d\gamma\left(t\right)/dt}\|\psi\|_2^2,
\]
where
\[
\|\psi\|_2^2=\int_{\Omega_q}|\psi|^2\sqrt{|g|}dq^1dq^2.
\]
Then,
\[
\frac{d}{dt}\left(\|\psi\|_2^2e^{-\gamma(t)}\right)=-\frac{Pe^{-1}}{\rho\left(t\right)}e^{-\gamma(t)}\|\mathrm{grad}\psi\|_2^2.
\]
Integrating,
\[
\rho_{\mathrm{min}}\int|\psi|^2\sqrt{|G|}dq^1dq^2\leq\|\psi\|_2^2\left(t\right)\leq \|\psi\|_2^2\left(0\right)e^{\gamma\left(t\right)}\leq\|\psi\|_2^2\left(0\right)e^{\gamma_0},
\]
since the growth function $\gamma\left(t\right)$ is bounded.  This gives
the required result.
\end{prop}

Before outlining the homogenization method, we re-work Eq.~\eqref{eq:deltac2} such
that we are left with the simplest possible equation.
First, by re-defining time, $\tau=\int dt\rho\left(t\right)^{-1}$, the
equation to homogenize has time-dependence only in the velocity and reaction
terms:
\[
\left[\frac{\partial c}{\partial \tau}-s\left(\tau\right)\right]\psi=\left[-\bm{V}\left(\bm{q},\tau\right)\cdot\nabla_q+Pe^{-1}\lapl_0\right]\psi,
\]
where
\[
s\left(\tau\right)=Da\rho\left(\tau\right)\left(1-2c_0\left(\tau\right)\right)-\frac{d\log\rho}{d\tau},\qquad
{V}^i=U^i\left(\bm{q},\tau\right)\rho\left(\tau\right).
\]
The time-dependent source term can be eliminated altogether through the use
of the equation
\begin{equation}
\frac{\partial\Psi}{\partial \tau}=\left[-\bm{V}\left(\bm{q},\tau\right)\cdot\nabla_q+Pe^{-1}\lapl_0\right]\Psi,
\label{eq:homogenize_simple}
\end{equation}
where
\[
\psi=\Psi\left(\bm{q},\tau\right)e^{\int s\left(\tau\right)d\tau}.
\]
Note that the transformation $\tau=\int^t ds/\rho\left(s\right)$ is well-defined
because it is invertible, $d\tau/dt=\rho\left(t\right)^{-1}\geq0$.
We now focus on homogenizing Eq.~\eqref{eq:homogenize_simple}).
 To do this, we must distinguish the time- and length scales in the problem.
  We work
 with a velocity field ${\bm{V}}\left(\bm{q},t\right)$ that varies rapidly
in space and time, on scales $\varepsilon$ and $\varepsilon^2$, respectively.
 The parameter $\varepsilon$ is obtained from the ratio $\varepsilon=\ell/L$,
 where $\varepsilon$ is the correlation length of the velocity field, and
 $L$ is the domain size.  We have chosen the timescale of variation to be
 $\varepsilon^2$ for definiteness, although other, slower and faster, scales
 are possible.  We assume that $\rho\left(t\right)$ is periodic in time and
 varies on the scale $\varepsilon^2$
 while and $G_{ij}\left(\bm{q}\right)$ varies only on the small length scale
 $\varepsilon$, or on the box-size scale, but not on both scales.

\paragraph*{The metric has small-scale variations:} 
We introduce the auxiliary function $\Psi_\varepsilon=\Psi\left(\qv/\varepsilon,\tau/\varepsilon^2\right)$,
which satisfies the equation
\begin{equation}
\frac{\partial\Psi_\varepsilon}{\partial \tau}=\left[-\varepsilon^{-1}\bm{V}\left(\bm{q},\tau\right)\cdot\nabla_q+Pe^{-1}\lapl_0\right]\Psi_\varepsilon.
\label{eq:pde_eps}
\end{equation}
Next, we introduce auxiliary independent variables $\bm{Q}=\qv/\varepsilon$ and $\sigma=\tau/\varepsilon^2$, such that
\begin{eqnarray*}
\grad &=& \grad_{\qv} + \varepsilon^{-1}\grad_{\Qv} \\
\lapl &=& \lapl_{\qv} + 2\varepsilon^{-1}\mathrm{div}_{\Qv}\mathrm{grad}_{\qv} + \varepsilon^{-2}\lapl_{\Qv},
\end{eqnarray*}
while the time derivative now also has two components:
\[
\frac{\partial}{\partial\tau}+\varepsilon^{-2}\frac{\partial}{\partial\sigma}.
\]
The PDE now reads,
\[
\left(\frac{\partial}{\partial\tau}+\varepsilon^{-2}\frac{\partial}{\partial\sigma}\right)\Psi_\varepsilon=\left(\varepsilon^{-2}\mathcal{L}_0+\varepsilon^{-1}\mathcal{L}_1+\mathcal{L}_2\right)\Psi_\varepsilon,
\]
where
\begin{eqnarray*}
\mathcal{L}_0&=&-V^i\left(\Qv,\sigma\right)\frac{\partial}{\partial
Q^i}+Pe^{-1}\lapl_{\Qv},\\
\mathcal{L}_1&=&-V^i\left(\Qv,\sigma\right)\frac{\partial}{\partial
q^i}+2Pe^{-1}\mathrm{div}_{\Qv}\mathrm{grad}_{\qv},\\
\mathcal{L}_2&=&Pe^{-1}\lapl_{\qv}.
\end{eqnarray*}
We expand the function $\Psi_\varepsilon$ in powers of
$\varepsilon$, as 
\[\Psi_\varepsilon(\Qv,\qv,\tau,\sigma)=
\Psi_0(\Qv,\qv,\tau,\sigma)+ \epsilon \Psi_1(\Qv,\qv,\tau,\sigma) + \epsilon^{2} \Psi_2(\Qv,\qv,\tau,\sigma)+....
\]
Equating powers of $\varepsilon$ in the expansion of the equation~\eqref{eq:pde_eps},
we obtain the following triad of problems:
\begin{eqnarray}
\frac{\partial\Psi_0}{\partial\sigma}-\mathcal{L}_0\Psi_0&=&0,\label{eq:triad1}\\
\frac{\partial\Psi_1}{\partial\sigma}-\mathcal{L}_0\Psi_1&=&\mathcal{L}_1\Psi_0\label{eq:triad2},\\
\frac{\partial\Psi_2}{\partial\sigma}+\frac{\partial\Psi_0}{\partial\tau}-\mathcal{L}_0\Psi_2&=&\left(\mathcal{L}_1\Psi_1+\mathcal{L}_2\Psi_0\right).\label{eq:triad3}
\end{eqnarray}
%
%
%
%
%
%
By multiplying Eq.~\eqref{eq:triad1} by $\sqrt{|G|}\left(\bm{Q}\right)\Psi_0\left(\bm{Q},\sigma\right)$ and integrating,
we obtain
\[
\|\Psi_0\|_2^2\left(\sigma_0\right)-\|\Psi_0\|_2^2\left(0\right)=-\int_0^{\sigma_0}\|\mathrm{grad}_Q\Psi_0\|_2^2d\sigma.
\]
Thus, if $\Psi_0$ is $\sigma_0$ periodic, then $\Psi_0$ is independent of $\Qv$,
and hence, is independent of $\sigma$.  In symbols,
\[
\Psi_0=\Psi_0\left(\qv,\tau\right).
\]
Using this solution, the second equation~\eqref{eq:triad2} becomes
\[
\frac{\partial\Psi_1}{\partial\sigma}-\mathcal{L}_0\Psi_1=-V^i\left(\bm{Q},\sigma\right)\frac{\partial\Psi_0\left(\bm{q},\tau\right)}{\partial
q^i}.
\]
This is solved by the ansatz
\[
\Psi_1=\theta^i\left(\bm{Q},\sigma\right)\frac{\partial\Psi_0}{\partial
q^i},
\]
where $\theta^i\left(\Qv,\sigma\right)$ solves the cell problem
\[
\frac{\partial\theta^i}{\partial\sigma}-\mathcal{L}_0\theta^i=-V^i,\qquad
\int_0^{\sigma_0}d\sigma\int_{\Omega_Q} \theta^i\left(\Qv,\sigma\right)\sqrt{|G|}dQ^1dQ^2=0,
\]
and where $\theta^i\left(\Qv,\sigma\right)$ is periodic in $\Qv$ and $\sigma$,
with periods $1$ and $\sigma_0$.
Finally, Eq.~\eqref{eq:triad3} has a solution provided
\[
\int_0^{\sigma_0}\int_{\Omega_Q}\left[-\frac{\partial\Psi_0}{\partial\tau}+\mathcal{L}_1\Psi_1+\mathcal{L}_2\Psi_0\right]\sqrt{|G|}dQ^1dQ^2=0,
\]
that is, if
\[
\frac{\partial\Psi_0}{\partial\tau}\left(\tau,\qv\right)=Pe^{-1}\lapl_q\Psi_0\left(\tau,\qv\right)+\frac{1}{|\Omega_Q|\sigma_0}\int_0^{\sigma_0}d\sigma\int_{\Omega_Q}\mathcal{L}_1\Psi_1\sqrt{|G|}dQ^1dQ^2.
\]
The second term can be re-arranged as
\begin{multline*}
\frac{1}{|\Omega_Q|\sigma_0}\int_0^{\sigma_0}d\sigma\int_{\Omega_Q}
\sqrt{|G|}\left(\Qv\right)dQ^1dQ^2\left[-V^{i}\theta^{j}+\frac{2Pe^{-1}}{\sqrt{|G|}}\left(\frac{\partial}{\partial
Q^k}\left(\sqrt{|G|}G^{ki}\theta^j\right)\right)\right]\frac{\partial}{\partial q^i}\frac{\partial}{\partial q^k}\Psi_0\left(\qv,\sigma\right)\\
=\frac{1}{|\Omega_Q|\sigma_0}\int_0^{\sigma_0}d\sigma\int_{\Omega_Q}
\sqrt{|G|}\left(\Qv\right)dQ^1dQ^2\left[-V^{i}\theta^{j}\right]\frac{\partial}{\partial q^i}\frac{\partial}{\partial q^k}\Psi_0\left(\qv,\sigma\right)\\
=\mathcal{J}^{ij}\frac{\partial}{\partial q^i}\frac{\partial}{\partial q^j}\Psi_0\left(\qv,\tau\right),
\end{multline*}
where the result on the second line is a consequence of incompressibility
and the periodic boundary conditions.  The homogenization matrix
\[
\mathcal{J}^{ij}=\frac{1}{|\Omega_Q|\sigma_0}\int_0^{\sigma_0}d\sigma\int_{\Omega_Q}
\sqrt{|G|}dQ^1dQ^2\left[-V^{i}\theta^{j}\right]
\]
is constant because $\sqrt{|G|}=\sqrt{|G|}\left(\Qv\right)$.
At lowest order, the equation for $\Psi_0\left(\bm{q},\tau\right)$ is
\begin{eqnarray}
\frac{\partial\Psi_0}{\partial\tau}&=&\left(Pe^{-1}\lapl_q+\mathcal{M}^{ij}\frac{\partial}{\partial
q^i}\frac{\partial}{\partial q^j}\right)\Psi_0,\nonumber\\
&=&\mathcal{K}\Psi_0,
\label{eq:homo1}
\end{eqnarray}
and thus the full solution (to Eq.~\eqref{eq:deltac2}) $\psi$ has a slowly-varying spatial component, multiplied by a rapidly-varying
temporal envelope:
\[
\psi\left(\bm{Q},\sigma,\tau\right)=\Psi_0\left(\bm{Q},\sigma\right)e^{\int s\left(\tau\right)d\tau}.
\]
We note the following result:
\begin{prop} [Regularity of the solution $\Psi_0\left(\qv,\tau\right)$]\label{prop:reg}
The operator $\mathcal{K}$ is negative in the following sense:
\[
\int_{\Omega_q}dq^1 dq^2\psi\mathcal{K}\psi\leq0,
\]
for any square-integrable function $\psi$,
and hence, the solution $\Psi_0$ to the equation~\eqref{eq:homo1} is uniformly
bounded in space and time.

The proof of this statement follows from a straightforward computation based
on the $\mathcal{J}$-operator:
\begin{multline*}
\int_{\Omega_q}dq^1 dq^2\psi\underbrace{\mathcal{J}^{ij}\frac{\partial}{\partial q^i}\frac{\partial}{\partial
q^j}}_{=\mathcal{J}}\psi\\=
-\int_0^{\sigma_0}\int_{\Omega_Q}dQ^1dQ^2\int_{\Omega_q}dq^1dq^2\sqrt{|G|}\left(\Qv\right)U^i\left(\Qv\right)\theta^j\left(\Qv\right)\psi\left(\qv\right)\frac{\partial}{\partial
q^i}\frac{\partial}{\partial q^j}\psi\left(\qv\right)\\
=\int_0^{\sigma_0}\int_{\Omega_Q}dQ^1dQ^2\int_{\Omega_q}dq^1dq^2\sqrt{|G|}\left(\Qv\right)U^i\left(\Qv\right)\theta^j\left(\Qv\right)w_i\left(\qv\right)w_j\left(\qv\right),\\
=\int_0^{\sigma_0}\int_{\Omega_Q}dQ^1dQ^2\int_{\Omega_q}dq^1dq^2\sqrt{|G|}\left(\Qv\right)\left(\bm{U}\cdot\bm{w}\right)\left(\bm{\theta}\cdot\bm{w}\right),
\end{multline*}
where $w_i=\partial\psi/\partial q^i$.
We introduce the quantity $\theta_w=\bm{w}\cdot\bm{\theta}$, which satisfies
$\left(\partial_\sigma-\mathcal{L}_0\right)\theta_w=-\bm{U}\cdot\bm{w}$.
 Then,
\begin{eqnarray*}
\int_{\Omega_q}dq^1 dq^2\psi\mathcal{J}\psi&=&
-\int_0^{\sigma_0}d\sigma\int_{\Omega_q}dq^1dq^2\int_{\Omega_Q}dQ^1dQ^2\sqrt{|G|}\left(\Qv\right)\left[\frac{\partial}{\partial\sigma}\theta_w^2-\theta_w\mathcal{L}_0\theta_w\right],\\
&=&-\int_0^\sigma d\sigma\int_{\Omega_q} dq^1dq^2\|\mathrm{grad}_Q\theta_w\|_2^2\leq0.
\end{eqnarray*}
\end{prop}
Since $\Psi_0$ is uniformly bounded in space and time ($\tau$), our homogenized solution
is bounded, and thus consistent with the boundedness result of Prop.~\ref{prop:bdd}, provided
\[
e^{\int s\left(\tau\right)d\tau}=e^{\int \left[s\left(t\right)/\rho\left(t\right)\right]dt}
\]
is bounded in time.  This is certainly the case, since the growth function
$\gamma\left(t\right)$ is bounded (Prop.~\ref{prop:growth}).

\paragraph*{The metric has domain-scale variations:} As before, the PDE to
homogeninze reads:
\[
\left(\frac{\partial}{\partial\tau}+\varepsilon^{-2}\frac{\partial}{\partial\sigma}\right)\Psi_\varepsilon=\left(\varepsilon^{-2}\mathcal{L}_0+\varepsilon^{-1}\mathcal{L}_1+\mathcal{L}_2\right)\Psi_\varepsilon,
\]
where
\begin{eqnarray*}
\mathcal{L}_0\phi&=&-\mathrm{div}_Q\left(\bm{V}\left(\Qv,\sigma\right)\phi\right)+Pe^{-1}\lapl_{\Qv}\phi,\\
\mathcal{L}_1\phi&=&-\mathrm{div}_q\left(\bm{V}\left(\Qv,\sigma\right)\phi\right)+2Pe^{-1}\mathrm{div}_{\Qv}\mathrm{grad}_{\qv}\phi,\\
\mathcal{L}_2\phi&=&Pe^{-1}\lapl_{\qv}\phi.
\end{eqnarray*}
The difference is that now $|G|=|G|\left(\qv\right)$, where $\qv$ is the
macroscopic scale.  This forces the velocity field $V^i$ to have the behaviour $V^i\left(\Qv,\qv,\sigma\right)=\tilde{V}^i\left(\Qv,\sigma\right)/\sqrt{|G|}\left(\qv\right)$,
by incompressibility.
%
%
%
%
%
%
%
%
%
The triad of problems~\eqref{eq:triad1}--\eqref{eq:triad3} is unchanged,
and the solution to Eq.~\eqref{eq:triad1} is again a function only of the
macroscopic scales:
 \[
\Psi_0=\Psi_0\left(\qv,\tau\right).
\]
Using this solution, the second equation~\eqref{eq:triad2} becomes
\begin{eqnarray*}
\frac{\partial\Psi_1}{\partial\sigma}-\mathcal{L}_0\Psi_1&=&-\mathrm{div}_q\left(\bm{V}\left(\Qv,\qv,\sigma\right)\Psi_0\left(\qv,\tau\right)\right)\\
&=&-\frac{\Psi_0}{\sqrt{|G|}}\frac{\partial}{\partial q^i}\left(\sqrt{|G|}V^i\right)-V^i\frac{\partial\Psi_0}{\partial
q^i}\\
&=&-\frac{\Psi_0}{\sqrt{|G|}}\frac{\partial}{\partial q^i}\left(\tilde{V}^i\left(\Qv,\sigma\right)\right)-\frac{\tilde{V}^i}{\sqrt{|G|}}\frac{\partial\Psi_0}{\partial
q^i},\\
&=&-\frac{\tilde{V}^i}{\sqrt{|G|}\left(\qv\right)}\frac{\partial\Psi_0}{\partial
q^i}.
\end{eqnarray*}
This is solved by the ansatz
\[
\Psi_1=\frac{\tilde{\theta^i}\left(\bm{Q},\qv,\sigma\right)}{\sqrt{|G|}}\frac{\partial\Psi_0}{\partial
q^i},
\]
where $\theta^i\left(\Qv,\qv,\sigma\right)$ solves the cell problem
\[
\frac{\partial\tilde{\theta}^i}{\partial\sigma}-\mathcal{L}_0\tilde{\theta}^i=-\tilde{V}^i,\qquad
\int_0^{\sigma_0}d\sigma\int_{\Omega_Q} \tilde{\theta}^i\left(\Qv,\qv,\sigma\right)dQ^1dQ^2=0,
\]
that is,
\[
\frac{\partial\tilde{\theta}^i}{\partial\sigma}+\frac{\tilde{V}^j}{\sqrt{|G|}}\frac{\partial\tilde{\theta}^i}{\partial
Q^j}-Pe^{-1}\lapl_Q\tilde{\theta}^i=-\tilde{V}^i,\qquad
\]
The solution $\tilde{\theta}^i$ must also satisfy the periodicity condition
in $\Qv$ and $\sigma$, with periods $1$ and $\sigma_0$, respectively.
Note that the macroscopic variable $\qv$ appears parametrically in the equation
for $\tilde{\theta}^i$.
%
%
Finally, Eq.~\eqref{eq:triad3} has a solution provided
\[
\int_0^{\sigma_0}\int_{\Omega_Q}\left[-\frac{\partial\Psi_0}{\partial\tau}+\mathcal{L}_1\Psi_1+\mathcal{L}_2\Psi_0\right]dQ^1dQ^2=0,
\]
that is, if
\[
\frac{\partial\Psi_0}{\partial\tau}\left(\tau,\qv\right)=Pe^{-1}\lapl_q\Psi_0\left(\tau,\qv\right)+\frac{1}{|\Omega_Q|\sigma_0}\int_0^{\sigma_0}d\sigma\int_{\Omega_Q}\mathcal{L}_1\Psi_1\left(\qv,Qv\right)dQ^1dQ^2.
\]
The second term can be re-arranged as
\begin{multline*}
\frac{1}{|\Omega_Q|\sigma_0}\int_0^{\sigma_0}d\sigma\int_{\Omega_Q}
dQ^1dQ^2\left[-\frac{\tilde{V}^{i}\tilde{\theta}^{j}}{\sqrt{|G|}\left(\qv\right)}\frac{\partial}{\partial q^i}\frac{1}{\sqrt{|G|}\left(\qv\right)}\frac{\partial}{\partial
q^j}\Psi_0\left(\qv,\sigma\right)-\frac{\tilde{V}^i}{\sqrt{|G|}\sqrt{|G|}}\frac{\partial\Psi_0}{\partial
q^j}\frac{\partial\tilde{\theta}^j}{\partial q^i}\right]\\
+
\frac{2Pe^{-1}}{|\Omega_Q|\sigma_0}\int_0^{\sigma_0}d\sigma\int_{\Omega_Q}
dQ^1dQ^2\frac{\partial}{\partial Q^i}G^{ij}\frac{\partial}{\partial q^j}\left[\frac{\tilde{\theta}^\ell}{\sqrt{|G|}}\frac{\partial\Psi_0}{\partial
q^\ell}\right]\\
=\left[\frac{1}{|\Omega_Q|\sigma_0}\int_0^{\sigma_0}d\sigma\int_{\Omega_Q}
dQ^1dQ^2\left(-\tilde{V}^{i}\tilde{\theta}^{j}\right)\right]\frac{1}{\sqrt{|G|}\left(\qv\right)}\frac{\partial}{\partial
q^j}\frac{1}{\sqrt{|G|}\left(\qv\right)}\frac{\partial}{\partial q^j}\Psi_0\left(\qv,\sigma\right)\\
-\frac{1}{|G|}\left[\frac{1}{|\Omega_Q|\sigma_0}\int_0^{\sigma_0}d\sigma\int_{\Omega_Q}
dQ^1dQ^2\left(\tilde{V}^i\frac{\partial\tilde{\theta}^j}{\partial q^i}\right)\right]\frac{\partial\Psi_0}{\partial
q^j}\\
=\frac{1}{\sqrt{|G|}}\frac{\partial}{\partial q^i}\frac{\mathcal{J}^{ij}}{\sqrt{|G|}}\frac{\partial}{\partial
q^j}\Psi_0\left(\qv,\sigma\right)
\end{multline*}
where the vanishing of the term on the second line is a consequence of the periodic boundary
conditions in $\Qv$.  Thus, we have obtained the homogenized matrix
\[
\mathcal{J}^{ij}\left(\qv\right)=\frac{1}{|\Omega_Q|\sigma_0}\int_0^{\sigma_0}d\sigma\int_{\Omega_Q}dQ^1dQ^2\left[-\tilde{V}^{i}\tilde{\theta}^{j}\left(\Qv,\qv,\sigma\right)\right],
\]
which determines the equation for $\Psi_0\left(\bm{q},\tau\right)$:
\begin{eqnarray}
\frac{\partial\Psi_0}{\partial\tau}&=&\left(Pe^{-1}\lapl_q+\frac{1}{\sqrt{|G|}}\frac{\partial}{\partial
q^j}\frac{\mathcal{J}^{ij}}{\sqrt{|G|}}\frac{\partial}{\partial q^j}\right)\Psi_0,\nonumber\\
&=&\mathcal{K}\Psi_0.
\label{eq:homo2}
\end{eqnarray}
As before, the full solution (to Eq.~\eqref{eq:deltac2}) $\psi$ has a slowly-varying
spatial component, multiplied by a rapidly-varying
temporal envelope:
\[
\psi\left(\bm{q},\sigma,\tau\right)=\Psi_0\left(\bm{q},\tau\right)e^{\int s\left(\tau\right)d\tau},
\]
where the temporal envelope can depend on $\sigma$ and $\tau$.
The diffusion operator is negative because
\[
\int_{\Omega_q}\sqrt{|G|}dq^1 dq^2\psi\mathcal{K}\psi\leq0,
\]
for any square-integrable function $\psi$.  This can be easily verified
by examination of the operator
\[
\mathcal{J}=\frac{1}{\sqrt{|G|}}\frac{\partial}{\partial
q^i}\frac{\mathcal{J}^{ij}}{\sqrt{|G|}}\frac{\partial}{\partial q^j},
\]
which satisfies the relation
\begin{eqnarray*}
\int \sqrt{|G|}dq^1dq^2\frac{1}{\sqrt{|G|}}\psi\frac{\partial}{\partial
q^i}\frac{\mathcal{J}^{ij}}{\sqrt{|G|}}\frac{\partial\psi}{\partial q^j}&=&\int dq^1dq^2\psi\frac{\partial}{\partial
q^i}\frac{\mathcal{J}^{ij}}{\sqrt{|G|}}\frac{\partial\psi}{\partial q^j},\\
&=&-\int dq^1dq^2\frac{\mathcal{J}^{ij}}{\sqrt{|G|}}\frac{\partial\psi}{\partial q^j}\frac{\partial\psi}{\partial q^j}.\\
\end{eqnarray*}
Using the trick demonstrated in Prop.~\ref{prop:reg}, the negativity
of $\mathcal{J}$, and hence $\mathcal{K}$ is established, and thus $\Psi_0$
is uniformly bounded in space and time ($\tau$ and $t$), consistent with
Prop.~\ref{prop:bdd}.

In conclusion, we have obtained effective diffusion operators for the concentration
$\Psi$ in limits when the substrate variation is either large or small in
spatial extent (Eqs.~\eqref{eq:homo1} and~\eqref{eq:homo2}).  A result for
a metric that varies on both scales is also obtainable from a combination
of these two approaches, provided the metric is separable, $|G|=G_1\left(\qv/\varepsilon\right)G_2\left(\qv\right)$.
 We now turn to the calculation of the diffusion constant for a given manifold.

\subsection{Shear flow on a modulating torus}
\label{subsec:torus}
We choose an isotropically time-varying torus as our manifold $\mathcal{M}=\mathbb{T}^{2}$,
where both the outer and inner radii vary with time according to the
constraint $\frac{R(t)}{r(t)}=a>1$. The strict inequality is to preserve
the topological character of the torus as a ring torus, thus preventing any
degeneration into a horn- or spindle-like surface.
Isotropic motion means that neither one of the toroidal
angle coordinates is time-dependant.  Moreover, either or both of the toroidal
radii can be regarded as
controlling the scale function $\rho(t)$.  
We introduce orthogonal co-ordinates $\vartheta-\varphi$, such that $\bm{x}\cdot\hat{\bm{z}}=r\sin\varphi$,
where $\bm{x}$ is the position vector and $\hat{\bm{z}}$ is the constant
unit vector in the $z$-direction.
 Thus, the co-ordinate $\varphi$ describes changes in angle around the
 minor circle.  The line element is then
\[
ds^2=r^2d\vartheta^2+\left(R+r\cos\vartheta\right)^2d\varphi^2,
\]
and thus the metric is diagonal:
\begin{equation}
g_{ij}=\begin{pmatrix}
1&& 0&& 0\\
0&& r^{2}&& 0\\
0&& 0&& (R+r\cos\vartheta)^{2}
\end{pmatrix}.
\label{eq:TorusMetric}
\end{equation}
Hence, $\sqrt{g}=r(R+r\cos\vartheta)=r^{2}(a+\cos\vartheta)$. 
We introduce the scale
function:
\begin{equation}
\rho(t):=r(t)^{2}=\frac{R(t)^{2}}{a^{2}}
\label{rhotorus}
\end{equation}
which is determined by the time variation of either radius.
Now the Laplacian on a torus takes the form
\[
\Delta =\frac{1}{r^2}\frac{\partial^2}{\partial\vartheta^2}
+\frac{1}{\left(R+r\cos\vartheta\right)^2}\frac{\partial^2}{\partial\varphi^2}
-\frac{\sin\vartheta}{r\left(R+r\cos\vartheta\right)}\frac{\partial}{\partial\vartheta},
\]
which after utilizing the relation \eqref{rhotorus} can be simply
written as
\begin{equation}
\Delta=\frac{1}{\rho}\Delta_0=
\frac{1}{\rho}\left[\frac{\partial^2}{\partial\vartheta^2}
+\frac{1}{\left(a+\cos\vartheta\right)^2}\frac{\partial^2}{\partial\varphi^2}
-\frac{\sin\vartheta}{a+\cos\vartheta}\frac{\partial}{\partial\vartheta}\right].
\end{equation}

We choose a simple shear flow that varies on the fast scale $\Qv=\left(\varepsilon\vartheta,\varepsilon\varphi\right)$.
 Specifically,
\[
\bm{V}=\frac{1}{\sqrt{|G|}}\left(0,b\left(Q^1\right)\right).
\]
The cell problem to solve is thus
\begin{eqnarray*}
\frac{\partial\theta^1}{\partial\sigma}+\frac{b}{\sqrt{|G|}}\frac{\partial\theta^1}{\partial Q^2}-Pe^{-1}\lapl_Q\theta^1&=&0,\\
\frac{\partial\theta^2}{\partial\sigma}+\frac{b}{\sqrt{|G|}}\frac{\partial\theta^2}{\partial Q^2}-Pe^{-1}\lapl_Q\theta^2&=&-b\left(Q^1\right).
\end{eqnarray*}
The solution is $\bm{\theta}=\left(0,\theta^2\left(Q^1\right)\right)$, where
\[
-D\frac{d^2\theta^2}{d\left(Q^1\right)^2}=b\left(Q^1\right).
\]
Thus, the matrix $\mathcal{J}^{ij}$ is constant and is equal to
\[
\mathcal{J}=\left(\begin{array}{cc}0&0\\0&\mathcal{J}^{22}\end{array}\right),\qquad\mathcal{J}^{22}=\int_0^1\left|\frac{d\theta^2}{dQ^1}\right|^2dQ^1=Pe\int_0^1\left|\frac{db}{dQ^1}\right|^2dQ^1
\]
Note that only the trivial solution is possible if $b$ is replaced with $\rho\left(\sigma\right)b\left(Q^2\right)$:
in this case, the time-periodicity forces us to choose the zero-solution,
and $\mathcal{J}=0$.  For the non-trivial case, the homogenized diffusion
equation is
\[
\frac{\partial\Psi_0}{\partial\tau}=Pe^{-1}\left[\frac{\partial^2}{\partial\vartheta^2}
+\frac{1+Pe\mathcal{J}^{22}}{\left(a+\cos\vartheta\right)^2}\frac{\partial^2}{\partial\varphi^2}
-\frac{\sin\vartheta}{a+\cos\vartheta}\frac{\partial}{\partial\vartheta}\right]\Psi_0,
\]
which can be solved using standard techniques for self-adjoint operators
on bounded domains.

\subsection{The extinction of the catalyst}
The reduction of the advection term provides a method for understanding
the extinction problem.  We want to know under what circumstances a small
initial concentration $c\left(\qv,0\right)=\delta\psi\left(\bm{q},0\right)$,
$\delta\ll1$
will go extinct.  The linearized homogenization theory just developed is
appropriate here.
The homogenized solution is
\[
\psi\left(\qv,t\right)=e^{Da\,t}\Psi_0\left(\qv,t\right),
\]
where $\Psi_0$ satisfies a diffusion equation
\[
\rho\left(t\right)\frac{\partial\Psi_0}{\partial t}=Pe^{-1}\mathcal{K}\Psi_0.
\]
Since $\mathcal{K}$ is negative and the manifold $\mathcal{M}$ is
smooth,
there is a complete set of eigenfunctions $\{A_\kappa\left(\qv\right)\}_{\kappa=0}^\infty$,
with corresponding eigenvalues $\{-\lambda_\kappa^2\}_{\kappa=0}^\infty$.
Thus, the solution is given
by the superposition
\[
\psi\left(\qv,t\right)=\frac{\rho\left(0\right)}{\rho\left(t\right)}\sum_{\kappa=0}^{\infty}C_\kappa
A_\kappa\left(\qv\right)e^{Da\,t-\lambda_\kappa^2Pe^{-1}\int_0^t\frac{ds}{\rho\left(s\right)}},
\]
where the $C_\kappa$'s are constant.
For a periodic modulation $\rho_{\mathrm{min}}\leq\rho\left(t\right)\leq\rho_{\mathrm{max}}$,
the amplitude decays to zero if
\[
\mathrm{min}_\kappa\lambda^2_\kappa > Da\,Pe\left(\frac{\rho_{\mathrm{max}}}{\rho\left(0\right)}\right).
\]
\begin{figure}[htb]
\centering\noindent
\includegraphics[width=0.7\textwidth]{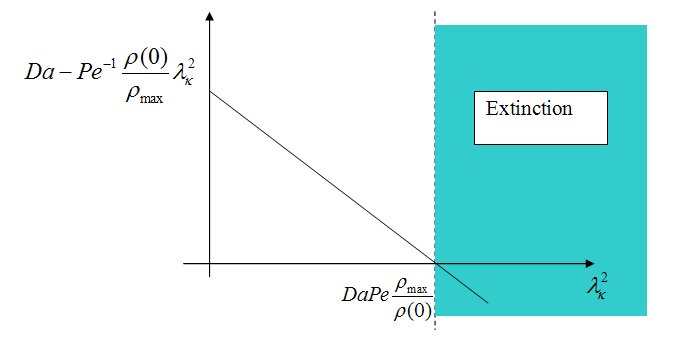}
\caption{A schematic diagram showing the possibility for the extinction of
the catalyst, when $\mathrm{min}_\kappa\lambda_\kappa^2>Da\,Pe\left[\rho_{\mathrm{max}}/\rho\left(0\right)\right]$.
 The time-varying scale factor $\rho_{\mathrm{max}}\geq\rho\left(0\right)$
 shifts the parameter range of extinction to the right, making extinction
 less likely.  On the other hand, the effective diffusion typically shifts
 the magnitude of the eigenvalues to larger values, thus moving the system
 further into, or closer to the domain of extinction.}
\label{fig:extinction}
\end{figure}
The parameter range of extinction is modified in two ways:
\begin{itemize}
\item The factor $\rho_{\mathrm{max}}\geq\rho\left(0\right)$ shifts the extinction
threshold to a higher value, as shown in Fig.~\ref{fig:extinction};
\item The contribution of the effective diffusion in Eqs.~\eqref{eq:homo1}
and~\eqref{eq:homo2} is negative
in sign,  and thus typically
increases the magnitude of the eigenvalues $\lambda_\kappa^2$ relative to
the unstirred value.  The minimum eigenvalue is then shifted rightward, thus
promoting extinction.
\end{itemize}
Having demonstrated how the full problem can be reduced to a diffusion-type
problem in a variety of different situations, we turn our attention to the
 numerical simulation of reaction-diffusion equations on various surfaces.

\section{Numerical case studies}
\label{sec:num}

In this section we examine three cases in which substrate modulation affects
the reaction propagation.  We focus on a modulating torus in three dimensions,
where the determinant of the metric tensor is separable, and on a
standing-wave substrate in two dimensions, in which case the determinant
of the metric tensor is not separable.  We also examine the bistable reaction
on the torus, in order to verify that our conclusion, namely that \textit{appropriate
substrate modulation enhances the reaction yield}, is independent of the
details of the reaction kinetics.

We use both analytical and numerical techniques.  Our numerical scheme is
a a semi-implicit spectral method in two dimensions.  In both situations,
the PDE to solve can be written as
\begin{equation}
\frac{\partial c}{\partial t}=\Delta c+\mu,\qquad
\mu=\mu\left(\frac{\partial^2 c}{\partial y^2},\frac{\partial c}{\partial
x},\frac{\partial c}{\partial y},c,|g|\right).
\label{eq:pde_mu}
\end{equation}
For the torus, we make the identification $x=\theta$, $y=\varphi$, while
for the substrate the variables $x$ and $y$ have their usual meaning.
We have also scaled all lengths in the problem according to an appropriate length scale $L$, and the time scale is taken to be $L^2/D$.  
For the toroidal case, $L=r_0$, a characteristic radius,
while for the substrate, $L$ is the periodic box size.
Given a time-periodic modulation, there are three non-dimensional frequencies
(timescales) in the problem: the diffusive frequency, here normalized to
unity, the modulation freqeuncy $\omega$, and the reaction frequency $\tilde{\sigma}=L^2\sigma/D$.
We are interested in cases where the effects of the chemical
reaction and the modulation greatly exceed the diffusive effects, and we
therefore take $\omega\approx \tilde{\sigma}\gg1$.  Following standard practice,
we henceforth omit ornamentation over non-dimensional quantities.
We Fourier-transform the PDE~\eqref{eq:pde_mu}, which takes the semi-implicit form
\[
\frac{\partial c_{\bm{k}}}{\partial t}\left(t_n\right)=-\bm{k}^2 c_{\bm{k}}\left(t_{n+1}\right)+\mu_{\bm{k}}\left(t_n\right),
\qquad
\mu_{\bm{k}}\left(t_n\right)=\int_{\left[0,L\right]^2}d^2\bm{x}e^{-i\bm{k}\cdot\bm{x}}\mu\left[\frac{\partial^2
c}{\partial y^2}\left(\bm{x},t_n\right),...\right],
\]
which can be integrated forward in time using standard techniques (a similar
approach has been used in solving the Cahn--Hilliard
equation; see~\cite{Zhu1999}.)  This method relieves the severe constraint
on the
timestep arising from a fully explicit treatment of the diffusion, and promotes
numerical stability.  Following standard practice, the discretization in
space and time is refined until convergence is achieved.

\subsection{A separable modulation on the torus}
\begin{figure}
\begin{center}
\subfigure[]{
\includegraphics[width=0.45\textwidth]{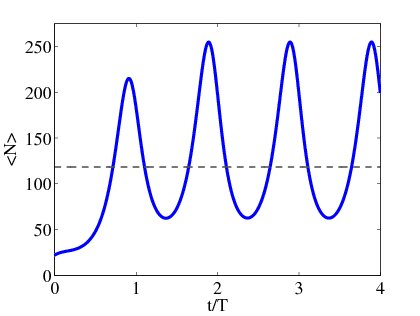}}
\subfigure[]{
\includegraphics[width=0.45\textwidth]{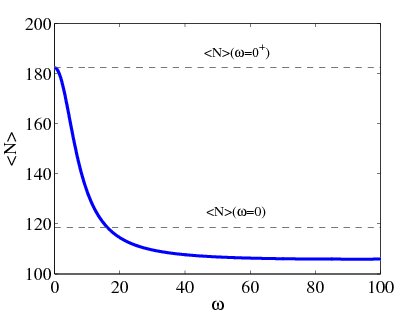}}
\end{center}
\caption{(a) The instantaneous yield $\langle N\rangle\left(t\right)$ for $\omega=10=\sigma$, and $\varepsilon=0.5$.
 The system settles down to a periodic state wherein the concentration
 fluctuates homogeneously.  The dashed line indicates the yield in the
 absence of pulsation; (b) The time-averaged yield as a function
 of the pulsation frequency.  The graph attains its
maximum as $\omega\rightarrow0$.  However, there is a discontinuity at $\omega=0$,
and the yield at zero frequency differs from that for very slow pulsations
$\omega\rightarrow0$.  This can be seen from Eqs.~\eqref{eq:mean_production}
and~\eqref{eq:mean_production1}  At large values of $\omega$, the yield
asymptotes to a constant value.  There is excellent agreement between the
yield values provided by this graph and the numerical solution of the PDE, and thus, the latter are not shown.} 
\label{fig:production} 
\end{figure}

We use the toroidal co-ordinate system outlined in Sec.~\ref{subsec:torus},
with the following radial modulation:
\begin{eqnarray}
R(t)&=& \frac{ar_0}{1+\varepsilon\sin\left(\omega t\right)}
\qquad0\leq\varepsilon<1,\nonumber\\
r\left(t \right)&=&\frac{R\left(t\right)}{a}, \qquad a>1,
\label{radiuspulsation_protocol}
\end{eqnarray}
with a  magnitude
$\varepsilon$ and constant angular frequency $\omega$, which gives
the pulsation protocol
\begin{equation}
\rho(t)=\frac{r_0^2}{\left[1+\varepsilon\sin\left(\omega t\right)\right]^{2}}.
\label{eq:growthpulsation_protocol}
\end{equation}
It should be noted that the torus, like the sphere, is special in the sense
that the scale function, which completely determines the geometric sink,
presents itself neatly in the form of the radius. Thus, this case can easily
be translated to a time-varying sphere.
The protocol~\eqref{eq:growthpulsation_protocol} modifies the yield of the reaction in the homogeneous
case.  The yield or production
is the time-average of the number of molecules created in the reaction, and
is defined by writing down the differential number of molecules in a a patch
of area
\[
dN\left(t\right)=c_0\left(t\right)dA\left(t\right).
\]
The integral of this quantity is the instantaneous yield:
\[
N\left(t\right)=4\pi^2a\rho\left(t\right)c_0\left(t\right),
\]
while the time-average of this quantity is the mean yield:
\[
\langle N \rangle =\lim_{t\rightarrow\infty}\frac{1}{T}\int_0^TN\left(t\right)dt.
\]
This mean yield is readily computable in the small-$\varepsilon$ limit.
 To do this, we use the pulsation relation~\eqref{radiuspulsation_protocol}, together with
the homogeneous solution~\eqref{bernoulli_sln}, to obtain the asymptotic
relation
%
%
%
%
\begin{multline}
\rho\left(t\right)c_0\left(t\right)\sim\\
\frac{r_0^2}{
1+\frac{2\sigma\omega\varepsilon}{\sigma^2+\omega^2}\left[\frac{\sigma}{\omega}\sin\left(\omega
t\right)-\cos\left(\omega t\right)\right]
+\tfrac{1}{2}\frac{\varepsilon^2}{\sigma^2+4\omega^2}\left[\sigma^2+4\omega^2-\sigma^2\cos\left(2\omega
t\right)-2\omega\sigma\sin\left(2\omega t\right)\right]
},\\
\text{as }t\rightarrow\infty.
\label{eq:c_longtime}
\end{multline}
This gives rise to the long-time average
\begin{multline}
\langle N\rangle=4\pi^2r_0^2a\;\lim_{T\rightarrow\infty}\frac{1}{T}\int_0^T\, N(t)dt,\\
=
{4\pi^2 r_0^2a}\lim_{T\rightarrow\infty}\frac{1}{T}\int_{0}^T
\frac{dt}{
1+\frac{2\varepsilon\sigma\omega}{\sigma^2+\omega^2}
\left(\frac{\sigma}{\omega}\sin\left(\omega{t}\right)
-\cos\left(\omega{t}\right)\right) +\tfrac{1}{2}{\varepsilon^2}
\left[1-\frac{\sigma^2\cos\left(2\omega{t}\right)+2\omega
\sigma\sin\left(2\omega{t}\right)}{\sigma^2+4\omega^2}\right]},\\
=4\pi^2r_0^2a \left[1+\tfrac{1}{2}\varepsilon^2
\frac{3\sigma^2-\omega^2}{\sigma^2+\omega^2}\right]
+O\left(\varepsilon^3\right),\qquad \omega> 0.
\label{eq:mean_production}
\end{multline}%
%
%
Thus, the amount of product created can either be raised or lowered, depending
on the reaction rate and the pulsation frequency; for large pulsations, the
yield is lowered.  The exact form of the yield $\langle N\rangle\left(\omega\right)$
is plotted in Fig.~\ref{fig:production}.
Note from this result that $\langle N\rangle\left(\omega\right)$ is not continuous at $\omega=0$.
 Setting $\omega$ to zero and then averaging gives $\langle N\rangle=4\pi^2r_0^2a$,
 which is the case of no pulsation.  On the other hand, for very slow pulsation
 (compared to the reaction rate),
 Eq.~\eqref{eq:mean_production} becomes
%
%
%
%
\begin{equation}
\langle N\rangle=
{4\pi^2 r_0^2 a}\lim_{T\rightarrow\infty}\frac{1}{T}\int_{0}^T
\frac{dt}{1+2\varepsilon\sin\left(\omega t\right)+\varepsilon^2\sin^2\left(\omega{t}\right)}
={4\pi^2 r_0^2 a}\left[\frac{\sqrt{1-\varepsilon^2}+2\varepsilon^2}{1-\varepsilon^2}\right].
\label{eq:mean_production1}
\end{equation}
Equation~\eqref{eq:mean_production1} is obtained by setting setting the terms
in Eq.~\eqref{eq:mean_production} to zero where $\omega$ appears as a factor.

%

The problem of front propagation on two-dimensional static manifolds has
been addressed by Gridndrod and Gomatam~\cite{Grindrod1987}, and thus some
qualitative details of front propagation are available.  Such analysis involves
the reduction of an equation in the Laplace--Beltrami operator to an associated
equation describing front propagation on the line.  Since the line is infinite
in extent, while the manifold in question is compact, this analysis is
valid only for intermediate times, after the front has been established,
but before the frontal region experiences the finite extent of the manifold.
We apply this technique to the torus by considering
an initial disturbance $c\left(\vartheta,\varphi,t=0\right)=c_0\left(\vartheta\right)$
centred on $\vartheta=\pi$.  This disturbance spreads only in the $\vartheta$
direction.
Thus $c\approx 1$ at $\vartheta\approx \pi$, while for early times, the reaction
has not propagated around the torus, and $c\approx0$ at $\vartheta\approx0$.
We denote the location of the front by $\vartheta_{\mathrm{f}}\left(t\right)$,
and introduce a moving
coordinate $\eta\left(\vartheta,t\right)=\vartheta-\vartheta_{\mathrm{f}}\left(t\right)$,
where the function $\vartheta_{\mathrm{f}}\left(t\right)$ is to be determined.
The profile of the front is given by a one-dimensional function: $c\left(\vartheta,t\right)=f\left(\eta\right)$,
and
\[
\frac{\partial{c}}{\partial t}=-f'\left(\eta\right)\frac{d\vartheta_{\mathrm{f}}}{dt}.
\]
Now for $\eta<0$, $f\approx1$, while for $\eta>0$, $f\approx0$.  Thus, we
are interested in a region $\eta\approx0$ where the profile of the function
$f$ changes rapidly.  We therefore write down the Laplacian in the neighbourhood
of this point:
\begin{eqnarray*}
\Delta c &=& \frac{\partial^2c}{\partial\vartheta^2}
-\frac{\sin\vartheta}{a+\cos\vartheta}\frac{\partial c}{\partial\vartheta},\qquad
a=R/r\\
%
%
%
%
%
%
&\approx&f''\left(\eta\right)
-\frac{\sin\vartheta_{\mathrm{f}}}{a+\cos\vartheta_{\mathrm{f}}}f'\left(\eta\right)
+O\left(\eta\right).
\end{eqnarray*}
Putting the reaction-diffusion equation together, we have
\[
f''\left(\eta\right)+f'\left(\eta\right)\left[\frac{d\vartheta_{\mathrm{f}}}{dt}-\frac{\sin\vartheta_{\mathrm{f}}}{a+\cos\vartheta_{\mathrm{f}}}\right]+\sigma
f\left(1-f\right)=0.
\]
If we stipulate the constant frontal velocity
\begin{equation}
k=\frac{d\vartheta_{\mathrm{f}}}{dt}
-\frac{\sin\vartheta_{\mathrm{f}}}{a+\cos\vartheta_{\mathrm{f}}}=\text{Const.},
\label{eq:stip}
\end{equation}
then we are reduced to the reaction-diffusion equation on the line,
for the variable $\eta$:
\[
f_{\eta\eta}+k f_\eta+\sigma f\left(1-f\right)=0.
\]
%
%
%
%
Eq.~\eqref{eq:stip} implies that that the velocity of the front is non-constant,
and evolves according to the differential equation
\begin{equation}
\frac{d\vartheta_{\mathrm{f}}}{dt}=k+\frac{\sin\vartheta_{\mathrm{f}}}{a+\cos\vartheta_{\mathrm{f}}}.
\label{eq:front_velocity}
\end{equation}
In addition to the constant term $k$, Eq.~\eqref{eq:front_velocity} possesses a curvature-related term that
can speed up or slow down the front propagation.
 In particular, there is the possibility of a stationary front when
\[
k+\frac{\sin\vartheta_{\mathrm{f}}}{a+\cos\vartheta_{\mathrm{f}}}=0.
\]
There is no analogue of the steady-state front in reaction-diffusion on the
line.  Here it
corresponds to a balance between the tendency of the reaction to propagate,
and the curvature of the torus, which inhibits the reaction propagation.
 For moderate to large $\sigma$-values $\sigma=10$--$1000$, the curvature
 term,
 being a diffusive contribution, is unimportant relative to the reaction
 term, and the dynamical equation for $\theta_{\mathrm{f}}$ gives approximately
 linear growth in time (this is verified by numerical simulation below).
Moreover, in this parameter range, it is possible to understand the modulated
solution by reference to a flat-space model.
Switching on the pulsation clearly will modify the front solution, since
the geometric
sink $-c\left(\partial\log\sqrt{g}/\partial t\right)=-c\rho'\left(t\right)/\rho\left(t\right)$
in the concentration equation breaks the Galilean
invariance.  However, some quantitative understanding of the front propagation
is still possible by studying the small-$\varepsilon$ equation
\begin{equation}
\frac{\partial c}{\partial t}=\left[1+\varepsilon d\left(t\right)\right]^2\Delta c + \sigma c\left(1-c\right)-\frac{2\varepsilon c d'\left(t\right)}{\left[1
+\varepsilon d\left(t\right)\right]^2},
\label{eq:rd_flat_front}
\end{equation}
where 
\[
\rho=\left[1+\varepsilon d\left(t\right)\right]^{-2},\qquad d\left(t\right)=\sin\left(\omega
t\right),
\]
and
\[
\Delta=\delta^{ij}\frac{\partial}{\partial x^i}\frac{\partial}{\partial x^j}.
\]
Using regular perturbation theory, it can be shown (Appendix~A) that the
location of the front in this case
is given by the formula
\[
\vartheta_{\mathrm{f}}\left(t\right)
=kt+\varepsilon\left[A\cos\left(\omega t\right)
+ B\sin\left(\omega t\right)\right]+O\left(\varepsilon^2\right),
\]
where $A$ and $B$ are amplitudes that are determined from the zeroth and
first-order solutions of the equation~\eqref{eq:rd_flat_front}, and $kt$ is
the reaction front when $\varepsilon=0$.  Note that
Mendez~\cite{Mendez2003} tackles a similar problem, but with slowly-varying
inhomogeneities and in flat space; the application we have in mind has rapidly-varying temporal
co-efficients.
Thus, the time-evolution for a periodic geometric sink is a secular
drift, coupled with a local-in-time back-and-forth oscillation as the function
$\rho\left(t\right)$ is modulated.  We can therefore give a qualitative description
of the front propagation on the modulating torus: there is a secular drift,
which is raised or lowered over the flat case due to curvature effects, while
there is a back-and-forth oscillation in the frontal position due to the
modulation of the toroidal area.  We turn to numerical simulations to check
this prediction.

\begin{figure}
\begin{center}
\subfigure[]{
\includegraphics[width=0.3\textwidth]{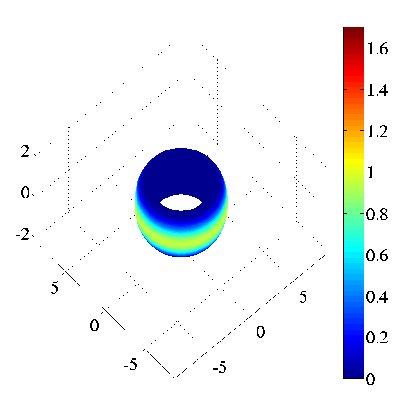}}
\subfigure[]{
\includegraphics[width=0.3\textwidth]{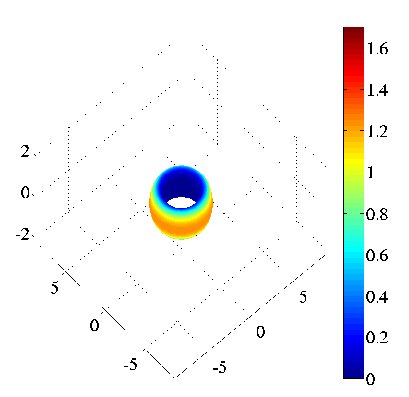}}
\subfigure[]{
\includegraphics[width=0.3\textwidth]{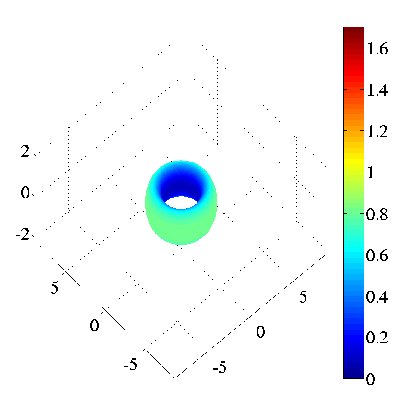}}
\subfigure[]{
\includegraphics[width=0.3\textwidth]{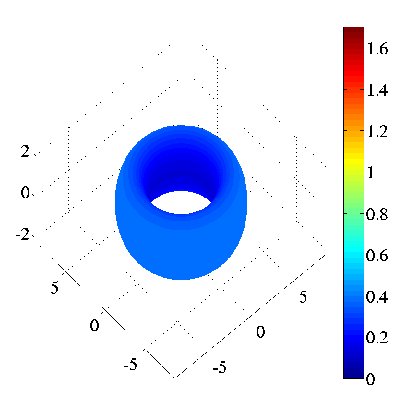}}
\subfigure[]{
\includegraphics[width=0.3\textwidth]{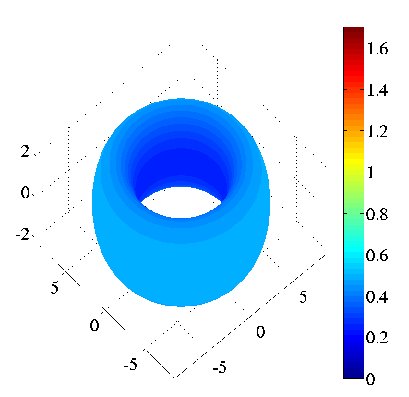}}
\subfigure[]{
\includegraphics[width=0.3\textwidth]{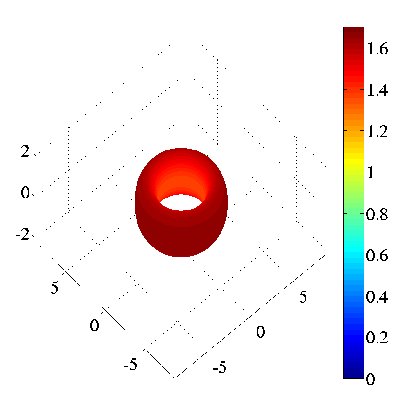}}
\end{center}
\caption{Front propagation on the torus at times $t=0$, $\tfrac{1}{5}T$,
$\tfrac{2}{5}T$, $\tfrac{3}{5}T$, $\tfrac{4}{5}T$, $T$, where $T=2\pi/\omega$ is the
period of the pulsation.  The catalyst is initially centred
on $\vartheta=\pi$ and propagates in both directions towards $\vartheta=0$.  The
concentration of catalyst tends to a uniform amount; however, as the torus
radii pulsate, the concentration level fluctuates, as shown in subfigures
(d)--(f).  We have taken $\omega=\sigma=10$, and $\varepsilon=0.5$.
} 
\label{fig:tori} 
\end{figure}

A numerical approach enables us to describe front propagation in the presence
of pulsation, and to verify the yield equation~\eqref{eq:mean_production}.
  We work with the pulsation protocol~\eqref{eq:growthpulsation_protocol}
and choose a ring-shaped
disturbance as an initial condition:
\[
c\left(\vartheta,\varphi,t=0\right)=e^{-\left(\vartheta-\pi\right)^2/2w^2},\qquad
w=0.2.
\]
Figure~\ref{fig:tori} shows the front propagation on the pulsating torus.
  The catalyst is initially centred
on $\vartheta=\pi$ and propagates in both directions towards $\vartheta=0$.  The
concentration of catalyst tends to a uniform amount; however, as the toroidal
radii pulsate, the concentration level fluctuates.
The most striking effect of the pulsating substrate is seen when the yield
of the reaction is studied, as in Fig.~\ref{fig:production}.  The yield fluctuates
over time.  The maximum yield exceeds the yield in the non-pulsating case,
while the minium yield lies below this steady value.  Fig.~\ref{fig:production}~(b)
shows the time-average yield as a function of pulsation frequency.  There
is a discontinuity at $\omega=0$, as discussed in the context of Eqs.~\eqref{eq:mean_production}
and~\eqref{eq:mean_production1}.
 At slow modulation frequencies, the average yield exceeds that of the non-pulsating
 case, while for faster modulation frequencies, the average yield decreases
 relative to this steady value.  Fig.~\ref{fig:production}~(b) also provides
 a verification of the yield formula Eq.~\eqref{eq:mean_production} and demonstrates
 the concention that surface modulation can enhance the yield. 
 
\begin{figure}
\begin{center}
\subfigure[]{
\includegraphics[width=0.45\textwidth]{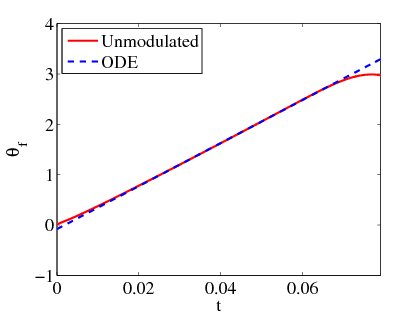}}
\subfigure[]{
\includegraphics[width=0.45\textwidth]{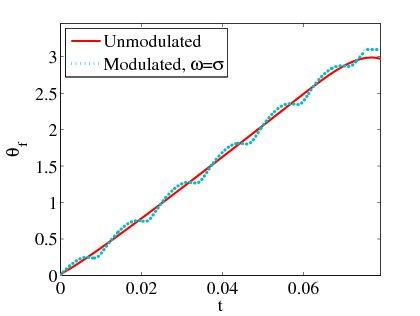}}
\end{center}
\caption{Front propagation on the torus, with $\sigma=500$.  Figure~(a) shows front propagation
on an unmodulated torus, and a comparison with the the front-tracking formula
$d\vartheta/dt=k+\left[\sin\vartheta/(a+\cos\vartheta)\right]$, with $k=43$. %
%
For
large times, the comparison is spoiled since the front wraps around the
torus.  Subfigure~(b) shows the modulated front.  The front drifts with the
same velocity as in the unmodulated case, while there is a backwards-and-forwards
motion as the toroidal area varies periodically.} 
\label{fig:fronts} 
\end{figure}

%
%
%
%
%
In the absence of pulsation, the speed of the front propagation satisfies
Eq.~\eqref{eq:front_velocity}, as confirmed in Fig.~\ref{fig:fronts}~(a).
When the pulsation
is switched on, there is still a net drift in the location of the front,
although locally in time, the front moves forwards and backwards as the surface
is modulated.  This confirmation of our earlier prediction is shown in Fig.~\ref{fig:fronts}~(b).
Next, we turn to the study of a qualitatively different case, that of
a standing-wave modulation on a substrate, in which case the determinant
\begin{figure}
\begin{center}
\subfigure[]{
\includegraphics[width=0.3\textwidth]{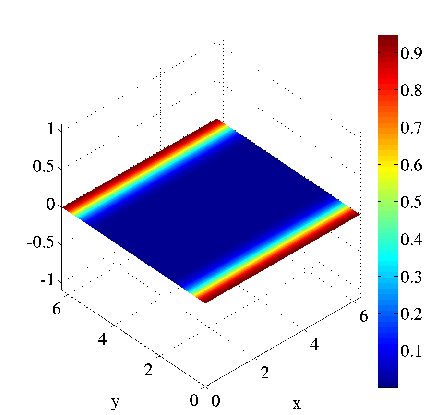}}
\subfigure[]{
\includegraphics[width=0.3\textwidth]{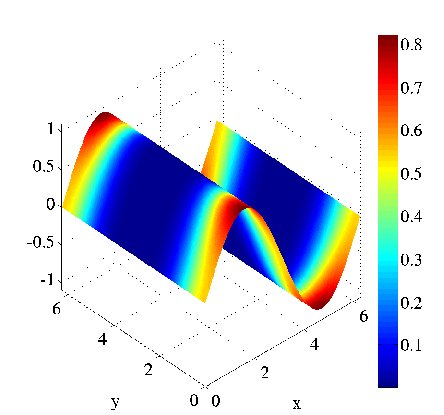}}
\subfigure[]{
\includegraphics[width=0.3\textwidth]{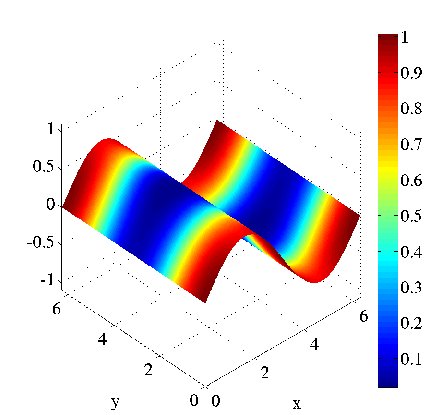}}
\subfigure[]{
\includegraphics[width=0.3\textwidth]{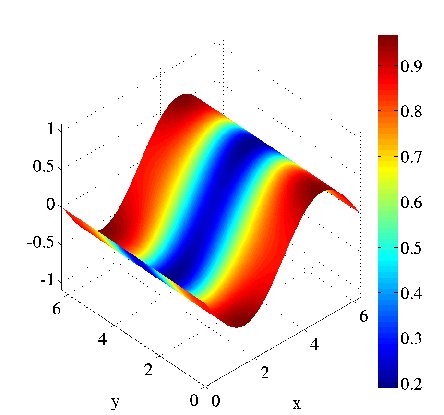}}
\subfigure[]{
\includegraphics[width=0.3\textwidth]{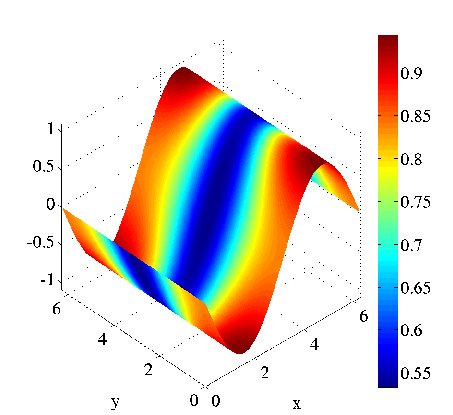}}
\subfigure[]{
\includegraphics[width=0.3\textwidth]{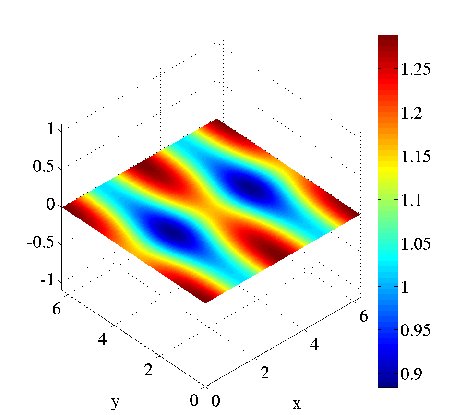}}
\end{center}
\caption{Front propagation on the substrate at times $t=0$, $\tfrac{1}{5}T$,
$\tfrac{2}{5}T$, $\tfrac{3}{5}T$, $\tfrac{4}{5}T$, $T$, where $T=2\pi/\omega$ is the
period of the pulsation, and $\omega=\sigma=10$.  The catalyst is initially centred
on $y=0$ and initially in the $y$-direction until a time-periodic state is reached.
 In reaching this state, the direction of the variation changes, as
 evidenced by subfigures (e) and (f).
} 
\label{fig:substrates1} 
\end{figure}
$|g|\left(x,y,t\right)$ is non-separable, and thus the geometric sink depends
on space and time.
 
\subsection{A non-seprable modulation: standing wave on a substrate}
In this section we work with a general periodic
surface embedded in
$\mathbb{R}^3$ with Cartesian coordinates $\left(x,y,z\right)$.
 The position vector $\bm{x}$ of a point $P\left(\bm{x}\right)$ on the substrate
 is given in the Monge parametrization as
\[
\bm{x}=\left(x,y,f\left(x,y,t\right)\right),
\]
where $f$ is a differentiable function of the planar coordinates $\left(x^1=x,x^2=y\right)$
and time.
The metric tensor is thus
\[
\left(g_{ij}\right)=\left(\begin{array}{cc}1+f_x^2&f_xf_y\\f_xf_y&1+f_y^2\end{array}\right),
\]
with inverse
\[
\left(g^{ij}\right)=\left(\begin{array}{cc}1+f_y^2&-f_xf_y\\-f_xf_y&1+f_x^2\end{array}\right).
\]
Both of these matrices have determinant
\[
|g|=1+f_x^2+f_y^2:=1+\left(\nabla_{\perp}f\right)^2.
\]
For a standing-wave surface
\[
f\left(x,y,t\right)=\varepsilon\sin\left(kx\right)\sin\left(\omega t\right),
\]
where $k$ and $\omega$ are constants, the Laplacian is
\[
\Delta=\frac{\partial^2}{\partial{x}^2}+\left(1+f_x^2\right)\frac{\partial^2}{\partial{y}^2}+\frac{f_xf_{xx}}{1+f_x^2}\frac{\partial}{\partial{x}},
\]
that is,
\[
\Delta=\frac{\partial^2}{\partial{x}^2}+\left[1+\varepsilon^2k^2\cos\left(kx\right)\sin\left(\omega{t}\right)\right]\frac{\partial^2}{\partial{y}^2}-\frac{\varepsilon^2k^3\sin\left(kx\right)\cos\left(kx\right)\sin^2\left(\omega{t}\right)}{1+\varepsilon^2k^2\cos^2\left(kx\right)\sin^2\left(\omega{t}\right)}\frac{\partial}{\partial{x}}.
\]
The chemical equation is thus given by
\begin{equation}
\frac{\partial c}{\partial t}=
\frac{\partial^2c}{\partial{x}^2}+\left(1+f_x^2\right)\frac{\partial^2c}{\partial{y}^2}+\frac{f_xf_{xx}}{1+f_x^2}\frac{\partial{c}}{\partial{x}}
+\sigma\left(c-c^2\right)
-\frac{2f_t f_{xt}}{1+f_x^2}c
\label{eq:c_substrate}
\end{equation}
\begin{figure}
\begin{center}
\subfigure[]{
\includegraphics[width=0.34\textwidth]{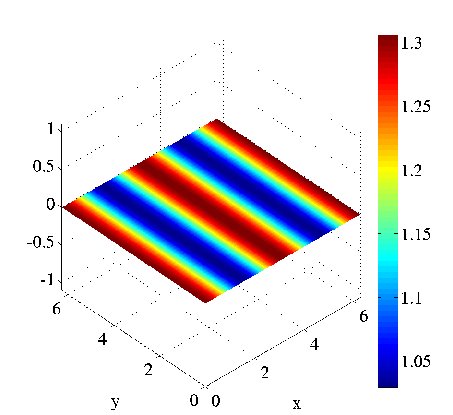}}
\subfigure[]{
\includegraphics[width=0.34\textwidth]{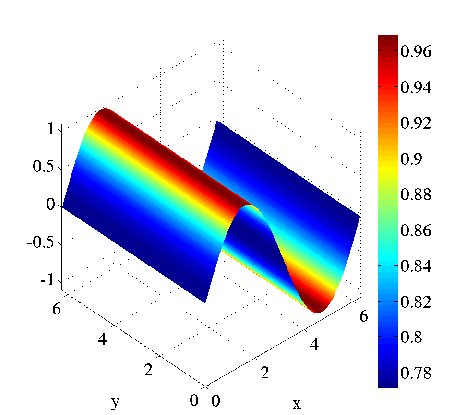}}
\subfigure[]{
\includegraphics[width=0.34\textwidth]{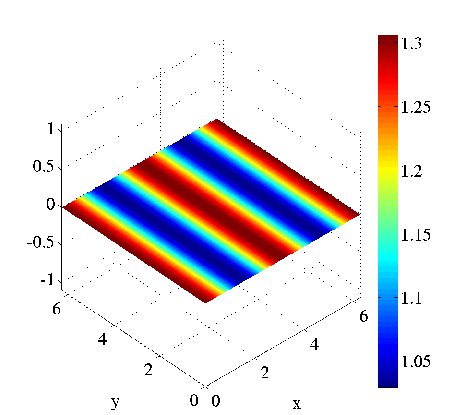}}
\subfigure[]{
\includegraphics[width=0.34\textwidth]{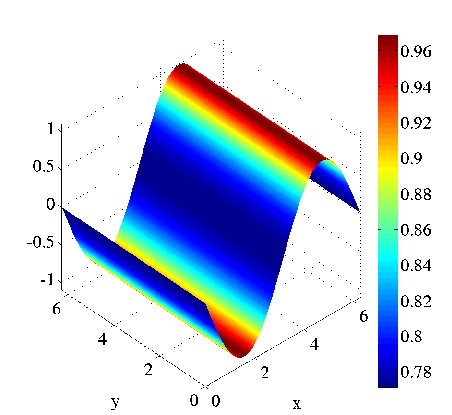}}
\end{center}
\caption{Front propagation on the substrate at times $t=3T$, $\tfrac{13}{4}T$,
and $\tfrac{14}{4}T$, and $\tfrac{15}{4}T$, where $T=2\pi/\omega$ is the
period of the pulsation, $\omega=\sigma=10$, $k=2\pi/L$, and $\varepsilon=1$.  The concentration has reached a time-periodic state
where all spatial variations are in the same direction as the direction of
the substrate modulation.} 
\label{fig:substrates2} 
\end{figure}
Figure~\ref{fig:substrates1} shows the case of front propagation for an initial
concentration level (Gaussian), with a wavenumber perpendicular to that of
the substrate modulation.
\begin{figure}
\begin{center}
\subfigure[]{
\includegraphics[width=0.45\textwidth]{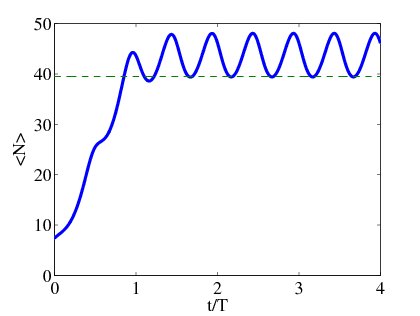}}
\subfigure[]{
\includegraphics[width=0.45\textwidth]{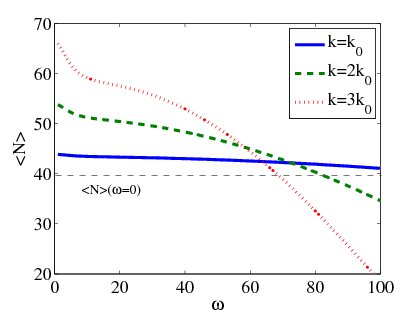}}
\end{center}
\caption{(a) The instantaneous yield $\langle N\rangle\left(t\right)$ for $\omega=10=\sigma$, $k=2\pi/\omega$, and $\varepsilon=1$.  This quantity is given
by the integral $\int dx\int dy\sqrt{1+f_x^2}c\left(x,y,t\right)$.
 The system settles down to a periodic state wherein the concentration
 flucuates inhomogeneously.  The dashed line indicates the yield in the
 absence of pulsation; (b) The average yield as a function of $\omega$
 and wavenumber $k$, where $k_0=2\pi/L$ is the fundamental wavenumber.  In
 general, for low pulsation frequencies, the yield is raised relative
 to the non-modulated case, while for fast frequencies, the yield is
 lowered.  The higher the wavenumber, the stronger the effect.}
\label{fig:production_substrate} 
\end{figure}
The front propagates
into regions of zero concentration, in an inhomogeneous fashion (since there
is spatial modulation in both directions).  After about one period of substrate
modulation, the spatial variation of the concentration field switches from
being in the $y$-direction, to being in the $x$-direction, aligned with the
substrate modulation. Eventually, the system attains a time-periodic state,
shown in Fig.~\ref{fig:substrates2}, where the dynamics are driven entirely
by the determinantal function $|g|\left(x,t\right)$.
On the other hand, for front propagation for an initial disturbance whose
wavenumber
is parallel to that of the substrate modulation, the f propagates into
regions of zero concentration in a homogeneous fashion, and the  system rapidly
reaches the time-periodic state shown in Fig.~\ref{fig:substrates2}.

The mean yield is always that associated with
with the time-periodic state, since any initial configuration tends asymptotically
to this state.  The yield function is
\[
\langle N\rangle\left(\omega,\varepsilon,k\right)=\bigg\langle \int dx \int dy \sqrt{1+f_x^2}c\left(x,y,t\right)\bigg\rangle.
\]
We obtain the yield function as a function of the parameters $\omega$, $k$,
and $\varepsilon$ by solving Eq.~\eqref{eq:c_substrate} numerically in one dimension
($\partial_y=0$).  The results
are shown in Fig.~\ref{fig:production_substrate}.  As before, the mean yield
as a function of time varies in phase with the substrate modulation, and
the maximum mean yield exceeds the stationary case.  The time-averaged mean
yield depends on the frequency of modulation: the slower the modulation,
the greater the yield.  Increasing the wavenumber of the modulation enhances
this effect, as seen in Fig.~\ref{fig:production_substrate}~(b).
In contrast with the toroidal case, the late-time state is not homogeneous,
 rather it varies periodically
 in space and time, according to the one-dimensional equation
\[
\frac{\partial c}{\partial t}=
\frac{\partial^2c}{\partial{x}^2}+\frac{f_x f_x^2}{1+f_x^2}\frac{\partial{c}}{\partial{x}}
+\sigma\left(c-c^2\right)
-\frac{2f_tf_{xt}}{1+f_x^2}c.
\]
An inhomogeneous final state is undesirable in applications where a pure
state involving only the product $B$ is required, and thus a pulsation protocol
similar to that on the torus is preferable over the substrate modulation
presented here.
 
\subsection{The bistable reaction on the torus}

We demonstrate numerically that the reaction yield can be enhanced for other,
more complicated mass-action laws, such as the bistable reaction.  Here,
there are two stable states $c=0$, and $c=1$, and an intermediate, unstable
state $c=\alpha_0$, where $0<\alpha_0<1$.  We study this reaction on the pulsating
torus; the relevant equation is
\begin{equation}
\frac{\partial c}{\partial t}=\Delta c + \sigma c\left(c-1\right)\left(\alpha_0-c\right)-c\frac{\partial\log\sqrt{g}}{\partial
t},\qquad \Delta=
\frac{1}{\rho}\left[\frac{\partial^2}{\partial\vartheta^2}
+\frac{1}{\left(a+\cos\vartheta\right)^2}\frac{\partial^2}{\partial\varphi^2}
-\frac{\sin\vartheta}{a+\cos\vartheta}\frac{\partial}{\partial\vartheta}\right],
\label{eq:bistable}
\end{equation}
where $\rho=r\left(t\right)^2$; for our pulsation protocol this is $\rho^{-1}=\left[1+\varepsilon\sin\left(\omega
t\right)\right]^2$.  

\begin{figure}
\begin{center}
\subfigure[]{
\includegraphics[width=0.45\textwidth]{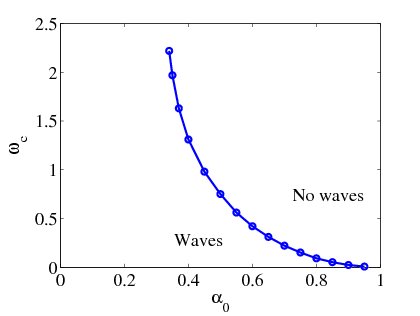}}
\subfigure[]{
\includegraphics[width=0.45\textwidth]{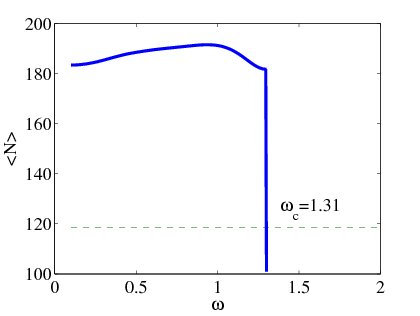}}
\end{center}
\caption{Characterization of the bistable reaction.  Subfigure~(a) gives
the parameter regimes in which either the zero state $c=0$, or the oscillatory
state, is selected as the asymptotic state.  The oscillatory state is preferred
at small frequencies, and the critical frequency is reduced at large $\alpha_0$-values.
 Subfigure~(b) gives the time-averaged yield as a function of $\omega$ for
 $\alpha_0=0.4$.  The time-averaged yield exceeds the stationary ($\omega=0$;
 dotted line)
 yield for $\omega<\omega_{\mathrm{c}}$, while for $\omega>\omega_{\mathrm{c}}$,
 the yield is zero.}
\label{fig:bistable} 
\end{figure}

Using a full two-dimensional numerical simulation, we
have verified that an arbitrary initial state tends either to the state $c=0$,
or a uniform oscillatory state.  The preferred state depends on the pulsation
parameters and the unstable level $\alpha_0$.  To see the relation between
these parameters, we studied the uniform equation, obtained by setting $\partial_\vartheta=\partial_\varphi=0$
in Eq.~\eqref{eq:bistable}.  We fixed $\varepsilon=0.5$ and $\sigma=10$ and
investigated the state selection as a function of $\omega$ and $\alpha_0$.
 For each value of $\alpha_0$ there is a critical frequency such that above
 that frequency, the zero state is preferred, while below that frequency,
 an oscillatory state is selected.  This relationship is shown in Fig.~\ref{fig:bistable}~(a).
  For large values of $\alpha_0$, close to $\alpha_0=1$, the critical frequency
  is shifted downward, indicating that the zero state is preferred for all
  but the slowest of modulation frequencies.  We have investigated the time-averaged
  mean yield as a function of $\omega$ and fixed $\alpha_0$.  Fig.~\ref{fig:bistable}~(b)
  shows this relationship for $\alpha_0=0.4$.  For $\omega<\omega_{\mathrm{c}}\left(\alpha_0=0.4\right)$,
  the time-averaged mean yield exceeds the stationary value (where $\omega=0$),
  while for $\omega>\omega_{\mathrm{c}}$ the mean yield is zero.  This result
  demonstrates that while more parameter-tuning is required, it is still
  possible to obtain a yield above the stationary yield simply by an appropriate
  modulation of the substrate.

\section{Conclusions}
\label{sec:conc}
 
We developed a mass-balance law for flow-driven chemical reactions on arbitary, time-varying
surfaces. The derivation is quite general, and takes into account situations
where the surface co-ordinates are themselves functions of time.  Our mass-balance
law possesses a geometric source / sink, which modifies the reaction.  For
isotropic surfaces, where the space- and time-dependence of the metric tensor
are separable, this geometric term is a function of time alone, and
a homogeneous solution is possible.  This solution is explicit
for the logistic reaction function, and the dependence of the concentration
level on the scale function of the metric tensor is thus made manifest.

In many situations~\cite{Plaza2004}, the surface of modulation is isotropic, and this case
therefore merits close attention.  We developed a theory for describing the
effects of flow for this class of manifold, and for flow fields with small-scale
spatial variations.  In such a scenario, homogenization theory permits one
to calculate the distribution of concentration through an effective-diffusion
equation.  Through surface modulation, the effective diffusion coefficient
depends on space, although this dependence is eliminated for a class of simple shear
flows on the torus; similar results for other surfaces are easily envisioned.

Having demonstrated a method for taking account of flow through the use of
an effective diffusivity, we focused on numerical simulations of reaction-diffusion
equations.  By numerically simulating logistic growth and diffusion on the
torus, we demonstrated the existence of reaction fronts
that drift at a constant velocity, but periodically advance and recede, due to
surface modulation.  We also demonstrated that the time-averaged yield of
the reaction could be increased by surface modulation.  A similar result
was found for the bistable growth law, although careful tuning of the modulation
frequency in relation to the bistable parameter is necessary for selection
of the required asymptotic state. For non-isotropic surfaces, the yield
was increased, although a spatially homogeneous state was impossible to attain.
 In summary, our PDE model and its simplifications provide an insight into
 the simultaneous processes of chemical reactions, stirring, and surface
 modulation, and should prove helpful in optimizing the outcome of chemical
 reactions on variable domains.

\subsection*{Acknowledgements}

The authors would like to thank G. Pavliotis for helpful suggestions.

%
\renewcommand{\theequation}{A-\arabic{equation}}
\renewcommand{\thefigure}{A-\arabic{figure}}
\setcounter{equation}{0}  
\setcounter{figure}{0}
\section*{APPENDIX A} 

In this section we calculate the perturbed speed of front propagation for
the following equation in flat space:
\begin{equation}
\frac{\partial c}{\partial t}=\left[1+\varepsilon d\left(t\right)\right]^2\Delta c + \sigma c\left(1-c\right)-\frac{2\varepsilon c d'\left(t\right)}{\left[1
+\varepsilon d\left(t\right)\right]^2},
\label{eqA:rd_flat_front}
\end{equation}
where 
\[
\rho=\left[1+\varepsilon d\left(t\right)\right]^{-2},\qquad d\left(t\right)=\sin\left(\omega
t\right),
\]
and
\[
\Delta=\delta^{ij}\frac{\partial}{\partial x^i}\frac{\partial}{\partial x^j}.
\]
The amplitude $\varepsilon$ is assumed to be small, $\varepsilon\ll 1$. 
We define the \textit{front} as the locus
of points $x_{\mathrm{f}}\left(t\right)$ in the uni-directional solution
$c\left(x,t\right)$, for which
\begin{equation}
c\left(x_{\mathrm{f}}\left(t\right),t\right)=\text{some constant}=C_0.
\label{eqA:front_def}
\end{equation}
When $\varepsilon=0$, the front is located at $x=x_{\mathrm{f}}^{(0)}\left(t\right)=kt$,
where $k$ is the constant velocity, which enters into the equation for the
front profile:
\[
\frac{d^2 \phi_0}{d\eta^2}+k\frac{d \phi_0}{d\eta}+\sigma \phi_0\left(1-\phi_0\right),\qquad
\eta=x-kt,
\]
where $\phi\left(-\infty\right)=0$ and $\phi\left(\infty\right)=1$.  We expand
the solution to the perturbed problem in powers of $\varepsilon$:
\[
c=\underbrace{c_0\left(x,t\right)}_{=\phi_0\left(\eta\right)}+\varepsilon c_1\left(x,t\right)+O\left(\varepsilon^2\right).
\]
The location of the front must change in order for the constraint~\eqref{eqA:front_def}
still to be satisfied:
\[
c\left(x_{\mathrm{f}}^{(0)}+\varepsilon x_{\mathrm{f}}^{(1)},t\right)=C_0.
\]
By Taylor expansion,
\begin{equation}
x_{\mathrm{f}}^{(1)}=-\frac{c_1\left(x_{\mathrm{f}}^{(0)},t\right)}{\frac{\partial
c_0}{\partial x}|_{x_{\mathrm{f}}^{(0)}}}.
\label{eqA:front_expansion}
\end{equation}
The first-order equation is
\[
\frac{\partial c_1}{\partial t}=\Delta c_1+\frac{\partial F}{\partial c}\bigg|_{c_0}c_1-2\left[d\left(t\right)\Delta
c_0+d'\left(t\right)c_0\right].
\]
We introduce new variables $c_1=\phi_1\left(\eta,t\right)$.  Thus,
\begin{eqnarray*}
\frac{\partial \phi_1}{\partial t}&=&\mathcal{L}_\eta \phi_1-2\left[d\left(t\right)\frac{d^2\phi_0}{d\eta^2}+d'\left(t\right)\phi_0\left(\eta\right)\right],\\
&=&\mathcal{L}_\eta \phi_1 -d\left(t\right)b_1\left(\eta\right)-d'\left(t\right)b_2\left(\eta\right),
\end{eqnarray*}
where
\[
\mathcal{L}_\eta=\frac{\partial^2}{\partial\eta^2}+v\frac{\partial}{\partial\eta}+\frac{\partial
F}{\partial c}\bigg|_{c_0\left(\eta\right)}.
\]
Since $d$ is periodic, we can write $d=\Re\left(\delta_0 e^{-i\omega t}\right)$
without loss of generality, and thus $\phi_1\left(\eta,t\right)=e^{-i\omega t}\phi_\omega\left(\eta\right)$, where $\phi_\omega\left(\eta\right)$ satisfies the equation
\[
\mathcal{L}_\eta\phi_\omega=-i\omega\phi_\omega+\delta_0 b_1\left(\eta\right)-i\omega\delta_0b_2\left(\eta\right).
\]
Thus,
\[
c_1\left(x_{\mathrm{f}}^{(0)},t\right)=\phi_\omega\left(0\right)e^{-i\omega
t}
\]
and hence Eq.~\eqref{eqA:front_expansion} becomes
\[
x_{\mathrm{f}}^{(1)}=-\frac{\Re\left(\phi_\omega\left(0\right)e^{-i\omega t}\right)}{\phi'\left(0\right)}.
\]
In other words,
\[
x_{\mathrm{f}}\left(t\right)
=x_{\mathrm{f}}^{(0)}\left(t\right)+\varepsilon\left[A\cos\left(\omega t\right)
+ B\sin\left(\omega t\right)\right]+O\left(\varepsilon^2\right),
\]
where $A$ and $B$ are constants,
as claimed in Sec.~\ref{sec:num}.

\end{document}